\documentclass{aa}  
\usepackage{xcolor}

\newcommand{\rev}[1]{#1}

\newcommand{\revvv}[1]{#1}
\usepackage{graphicx}
\usepackage{amsmath}
\usepackage{subcaption}
\usepackage{mathtools} 
\usepackage{hyperref}
\usepackage{multirow}
\usepackage{tabularx}
\usepackage{txfonts}
\usepackage{booktabs}
\usepackage{longtable}
\usepackage[table]{xcolor}
\definecolor{lightgray}{gray}{0.93}

\begin{document}

   \title{DEMOS. II. What the full Lupus demographics truly tell us about protoplanetary disc evolution}

    \author{Lorenzo A. Malanga
          \inst{1}\thanks{\email lorenzo.malanga@unimi.it}
          \and
          Giovanni P. Rosotti
          \inst{1}
          \and
          Giuseppe Lodato
          \inst{1}
          \and
          Alice Somigliana
          \inst{2}
          \and
          Benoît Tabone
          \inst{3}
          \and 
          Rossella Anania
          \inst{4,1}
          \and
          Carlo F. Manara
          \inst{5}
          \and
          Leonardo Testi
          \inst{6,7}
          \and
          Claudia Toci
          \inst{8}
          \and
          Francesco Zagaria
          \inst{2}
          }

   \institute{Dipartimento di Fisica 'Aldo Pontremoli', Università degli Studi di Milano, Via Celoria 16, I-20133 Milano, Italy
             \and 
             Max-Planck Institute for Astronomy (MPIA), Königstuhl 17, 69117 Heidelberg, Germany
             \and 
             Université Paris-Saclay, CNRS, Institut d’Astrophysique Spatiale, 91405 Orsay, France
             \and
             School of Physics, Trinity College Dublin, University of Dublin, College Green, Dublin 2, Ireland
             \and
             European Southern Observatory, Karl-Schwarzschild-Strasse 2, D-85748 Garching bei München, Germany
             \and
             Dipartimento di Fisica e Astronomia, Università di Bologna, Via Gobetti 93/2, I-40122 Bologna, Italy.
             \and 
              INAF-Osservatorio Astrofisico di Arcetri, Largo E. Fermi 5, I-50125 Firenze, Italy.
              \and
              Escuela Técnica Superior de Ingeniería, Universidad de Sevilla, Camino de los Descubrimientos s/n, 41092 Sevilla, Spain
             \\
             }

   \date{Received..., 2026; accepted ..., ...}

 \abstract{Assessing whether disc evolution is driven by viscosity or Magneto Hydro-dynamic (MHD) winds is crucial to determine the conditions in which planets form.
 The rich statistical information encoded in the nearby star-forming regions provides an excellent opportunity to probe disc evolution.
 In this paper, we apply the Bayesian inference tool \texttt{DEMOS}, which is designed to fully exploit the statistical properties of disc populations and infer initial conditions and model parameters, to the Lupus disc population.
 We consider the observed distribution of stellar mass $M_\star$, disc mass $M$ (obtained from continuum flux), mass accretion rate $\dot{M}$ and CO disc size $R_{90\%, \text{CO}}$.
 We constrain all parameters for both the viscous model, which includes dispersal driven by internal photoevaporation, and the MHD winds model, with unprecedented precision.
 The inferred initial conditions suggest that positive correlations between initial disc properties, such as mass $M_0$ and truncation radius $R_0$, with the stellar mass $M_\star$, must be already present in the population before the disc phase.
 According to both models, the typical $R_0$, for a disc around a $0.3\ \text{M}_\odot$ star, is $3-5$ au, with a strong dependence on stellar mass of $\sim 2$ and $\sim 1$ for the viscous-photoevaporative and MHD winds models, respectively.
 We infer values for the angular momentum transport parameter $\alpha$ (or $\alpha_\text{DW}$) of $3-4 \times 10^{-4}$, in both models, for discs around $0.3\ \text{M}_\odot$ stars, and an approximately quadratic dependence on $M_\star$.
 For the viscous-photoevaporative model, we infer a typical wind's mass-loss rate $\dot{M}_\text{w} \sim 10^{-9}$ M$_\odot$ yr$^{-1}$ for a disc around a $0.3\ \text{M}_\odot$ star, and a steep correlation with $M_\star$ with an exponent $\sim 2.5$.
 The inferred mean age of the Lupus population differs between models, with $\langle t\rangle$ = 1.6 Myr for the MHD winds model and $\langle t\rangle$ = 0.9 Myr for the viscous-photoevaporative framework.
 Despite indicating a slight preference for the viscous-photoevaporative framework, the Bayesian evidence computed with \texttt{DEMOS} shows that both models reproduce the data with comparable accuracy and does not allow either scenario to be ruled out.
 However, the best-fit synthetic populations predict substantially different evolutionary pathways, suggesting that the application of \texttt{DEMOS} to multiple populations, especially to evolved ones such as Upper Scorpius, will allow to place stronger constraints on disc evolution mechanisms.
 
}

               
\titlerunning{Application of DEMOS to Lupus}
\authorrunning{Malanga, L. et al 2026}

\maketitle

\section{Introduction}
\label{introduction}

Protoplanetary discs are the environment where planet formation occurs, thus understanding the processes that govern their evolution and dispersal is paramount to explain the vast number of observed exoplanets.
Pioneering surveys have revealed the properties of protoplantetary discs across various star-forming regions. Among the most notable findings are ongoing accretion \citep{Hartmann2016}, with accretion rates that scale with stellar mass and decrease over time \citep{Herczeg2008, Alcala2014, Testi2022}, and a declining fraction of disc-bearing sources as a function of the population’s age (\rev{\citealt{Haisch2001}};\citealt{Hernandez2007, Fedele2010, Ribas2015}), with a typical dispersal time of 2-3 Myr.
Such properties can be explained by two distinct mechanisms.
The viscous paradigm \citep{ShakuraSunyaev1973} asserts that angular momentum is transported throughout the disc thanks to turbulence, inducing the disc to expand with time \citep{LyndenBellPringle1974}.
In this scenario, the fast disc dispersal is attributed to photoevaporative winds launched from the disc surface when heated by UV and X-ray radiation from the host star \citep{Hollenbach1994, Alexander2014, Nakatani2018, Picogna2019, Sellek2024b}.
A combination of turbulence-driven accretion and photoevaporative clearing reproduces both the observed accretion rates and typical dispersal timescales (\citealt{Clarke2001, Ruden2004, Alexander2006b, Owen2012, Somigliana2020};\rev{\citealt{Emsenhuber2023}}).
On the other hand, both accretion and dispersal can be explained by the presence of magneto-hydrodynamic (MHD) winds, which remove both mass and angular momentum from the disc \citep{BlandfordPayne1982, Armitage2013}.
The development of 1D solutions for the MHD winds model \citep{Suzuki2016, Tabone2022model} has allowed to compare the predicted evolution of global disc properties to the viscous-photoevaporative framework on a population level.

With the advent of a new generation of telescopes and interferometers, numerous surveys have measured global properties of protoplanetary discs such as disc masses \citep{Ansdell2016, Ansdell2017, Barenfeld2016, Miotello2016, Pascucci2016, Bergin2017, Cazzoletti2019, Grant2021, Testi2022, Carpenter2025}, accretion rates \citep{Alcala2014, Alcala2017, Manara_2017, Venuti2019, Manara2020, Fang2023, Almendros-Abad2024, Empey2026}, and disc sizes \citep{Tazzari2017,   Tazzari2021,Andrews2018, Ansdell2018, Najita2018,Hendler2020, Sanchis2020, Sanchis2021, Zallio2026_usco, Ricciardi2026} across a range of star-forming region.

The rich statistical samples made available by these surveys offer an exquisite benchmark to test disc evolutionary models (see \citealt{Manara2023} for a review).
Over the past decade, several studies proposed various methods to exploit the statistical properties of key observables to distinguish between the viscous framework and the wind-driven scenario (\citealt{Lodato2017, Mulders2017, Rosotti2017, Rosotti2019, Najita2018, Trapman2020, Somigliana2022, Somigliana2023, Somigliana2024, Alexander2023, Zallio2024, Malanga2025}; \rev{\citealt{Weder2026}}).
However, neither of these evolutionary models has yet been definitively ruled out.
A significant milestone in disc population synthesis is represented by \cite{Tabone2025}, who used the 1D population synthesis code \texttt{Diskpop}, presented by \cite{Somigliana2024}, to reproduce the AGE-PRO subsamples of Lupus, Ophiucus and Upper Scorpius \citep{Zhang2025, Deng2025}.
They developed a systematic model to constrain four key parameters of both the viscous and MHD winds model, by jointly fitting four observables: the median disc mass, the median accretion rate, the median disc size and the disc fraction.
Their results show that the viscous model fails to simultaneously reproduce disc masses and accretion rates, pointing instead to a preference for the MHD winds model.
However, this study did not exploit the full potential of the Lupus demographics;
in fact, the AGE-PRO large program included only a limited subsample of Lupus and within a narrow range of stellar masses.
Moreover, fitting only for median observed values, they were unable to test more complex evolutionary models, such as those predicting correlation between initial disc properties and the stellar mass \citep{Alexander_Armitage2006, Dullemond2006} nor to consider the information encoded in the spread of the distributions \citep{Lodato2017, Somigliana2023}.
This limitation highlights the need for an automated framework capable of exploring vast parameter spaces. 
Such a procedure becomes essential as soon as we consider models that incorporate environmental effects, such as the infall of material from the interstellar medium onto accreting discs, known as late infall \citep{Ginski2021, Gupta2023, Huang2024, Winter2024, Speedie2025} or ultraviolet radiation emitted by nearby massive stars, which can drive external photoevaporation \citep{Clarke2007, Haworth2019, Winter2022, Anania2025AGE-PRO, Pichierri2026}.
Moreover, \cite{Tabone2025} relied on a the step-by-step fitting procedure that could not rigorously quantify the uncertainties of constrained parameters.

To address this problem, we developed \texttt{DEMOS}, a Bayesian inference tool designed to constrain model parameters by fully leveraging the demographic properties of disc populations with statistically rigorous techniques.
We presented \texttt{DEMOS} in Malanga et al. (2026a) (hereafter Paper I), where we describe its methodology and demonstrate its strengths.
In this study, we apply the tool to the Lupus population, with the goal to determine the model parameters and initial conditions that best reproduce the observations.
This is the first time, in the field of protoplanetary discs, in which statistically robust techniques are used to exploit the full demographics of a disc population and simultaneously infer model parameters.
This paper is structured as follows.
In Section \ref{observations} we describe the Lupus population and the criteria that we used to select the sample for our analysis.
In Section \ref{model} we describe the tested evolutionary models and the free parameters of our inference.
In Section \ref{results} we illustrate the results that we obtain from the inference, emphasising the best-fit model parameters, the overall model accuracy and the difference between the disc evolution predicted by the two models.
Finally, in Section \ref{discussion} we discuss the achievements of our analysis, the possible implications of our results and future applications.

\section{The observed sample}
\label{observations}

Due to relative short distance of approximately 160 pc from our Solar System \citep{Lombardi2008}, the Lupus star-forming region is among the most frequently targeted to investigate planet formation.
Surveys carried out with \textit{Spitzer} \citep{Evans2009, Dunham2014} and later \textit{Gaia} \citep{Luhman2020} identified pre-main-sequence objects in the regions, with an exceptionally high degree of completeness ($\sim 80-90\%$) and covering the full range of stellar masses.
For this reason, Lupus represents the optimal target for population synthesis studies that rely on sample completeness, such as this work.
We note that the sample is almost complete in terms of dust masses \citep{Ansdell2016}, but only a small subsample is currently available with gas masses ($\sim 10\%$) \citep{Trapman2025_AGE_PRO}.
Even if the former suffer from significant uncertainties related to opacities, optical depth and dust-to-gas ratio \citep{Miotello2022}, we opt to employ them as we focus here on harnessing the full statistical power of the data.
This allows us to perform the inference without considering selection and observational biases, which would reduce the accuracy of the fit and are not currently implemented in \texttt{DEMOS} (see Paper I).
Moreover, the absence of massive O and B-type stars in the region makes Lupus a low-irradiation environment, with an estimated average FUV flux below 10 G$_0$ \citep{Anania2025}.
Therefore, it is suitable for testing models that do not include the effect of external photoevaporation.
The age of Lupus is estimated to be between 1 and 3 Myr based on its distribution in the Hertzsprung–Russell (HR) diagram \citep{Alcala2017, Galli2020_Lupus}; this estimate places the region in an intermediate evolutionary stage, older than notably young regions, such as Ophiucus \citep{Esplin2020}, and younger than Upper Scorpius \citep{Pecaut2016, Luhman&Esplin2020}.

\subsection{Disc masses and accretion rates}

We consider the Lupus sample compiled in the PPVII review \cite{Manara2023}, which encompasses 100 disc-bearing sources whose membership to the cluster has been confirmed by \textit{Gaia} \citep{Luhman2020}.
The sample includes measurements of the stellar mass $M_\star$, the dust mass derived from the continuum emission $M_\text{dust}$ and the mass accretion rate $\dot{M}$.
Since the evolution of discs in binary systems is not currently implemented in \texttt{Diskpop}, we focus on reproducing the demographics of the Lupus discs in single-star systems.
Therefore, we excluded from our sample 11 objects, which are confirmed multiple systems with projected separation $a$ lower than 300 au (see \citealt{Zurlo2021, Zagaria2022binaries}).
This choice is motivated by both theoretical \citep{Zagaria2021} and empirical \citep{Harris2012} results indicating that discs in binary systems with $a > 300$ au do not exhibit significant differences in their bulk properties from those evolving in isolation.
Out of the remaining 89 objects, $M_\star$ has been measured for 65 of them by \cite{Alcala2014, Alcala2017}, obtained using the evolutionary tracks of Pre-Main-Sequence (PMS) stars \citep{Siess2000, Baraffe2015, Feiden2016}.
Within this subset, \cite{Alcala2014, Alcala2017, Alcala2019} measured the mass accretion rate for 62 sources, using the UV-excess observed by the Very Large Telescope/ X-Shooter instrument. 
They further reported an upper limit on $\dot{M}$ for 3 sources due to non-detections.
For our analysis, we consider the $\dot{M}$ sample of \cite{Manara2023}, who revised the original values using the more accurate distance measurements from \textit{Gaia}.
For objects where $\dot{M}$ or $M_\star$ has not been measured, we adopted broad, non-informative upper limits of $10^{-7} \ \text{M}_\odot$ yr$^{-1}$ and $3 \ \text{M}_\odot$, respectively.
These upper limits are not intended to provide a value for those quantities but rather to express our lack of information and allow the likelihood (see Section 2.2 of Paper I) to account for it appropriately. 
Dust masses are derived from continuum emission, observed by ALMA surveys \citep{Ansdell2016, Ansdell2018, Sanchis2020}, for 65 sources, while \cite{Ansdell2016} reported upper limits for the remaining 24 sources.
These measurements rely on the assumption of optically thin emission and known dust opacities, with caveats discussed in the review by \cite{Miotello2022}.
We convert dust masses $M_\text{dust}$ to gas masses $M$ assuming a gas-to-dust ratio of 100, consistent with the value from the ISM.

\subsection{Disc sizes}
\label{disc_sizes}

The size of discs has been estimated with both dust and gas tracers. 
However, dust evolution processes, such as radial drift, can decouple the dust distribution from the gas, limiting the reliability of dust emission as a proxy for the total disc extent \citep{Rosotti2019, Toci2021, Zagaria2022}. 
Consequently, we consider the measurements of the disc radius $R$ derived from the $^{12}$CO $ J = 2 - 1$ rotational transition from \cite{Sanchis2021} and \cite{Trapman2025_AGE_PRO_sizes} and from $^{12}$CO $ J = 3 - 2$ from \cite{Ricciardi2026}.
We considered sizes measurements from \cite{Trapman2025_AGE_PRO_sizes} for the 8 resolved sources in common to both our sample and AGE-PRO.
We further considered \cite{Ricciardi2026} for 13 of the discs in our sample which are not included in AGE-PRO, out of which 8 are detections and 5 are upper limits.
\cite{Sanchis2021} estimated the radius that encloses 68\% of the flux for 35 discs within our sample, from which we selected the 29 sources that are not included in the two above mentioned surveys. 
In addition, they provided upper limits for EX Lup and V1192 Sco because of reported outflow that may contaminate the emission, and of SONYC-Lup3-7 and Lup706 due to large uncertainties associated with systems with very low mass stars.
For three sources, namely J15450634-3417378, J15450887-3417333 and J16011549-4152351, we adopted the non-informative upper limit of 10000 au, because their emission is partially covered by the cloud.
For the remaining 31 sources in our sample, for which no survey provided a radius estimate, we derived values using the flux-radius relation of the Lupus population \citep{Zagaria2023b}, calibrated on the detections of \cite{Sanchis2021}. 
We adopted flux measurements for detections and upper limits for non-detections from \cite{Miotello2021}, from which we derived either a radius estimate or an upper limit using this relation.
This procedure is justified by the result of \cite{Miotello2021}, and more recently \cite{Ricciardi2026}, that faint or undetected CO emissions are compatible with compact discs models.
We used the same criterion to estimate the size of Sz66, for which \cite{Sanchis2021} provide a consistent upper limit due to its point-like nature.
\revvv{For Sz98, we put a non-informative upper limit since the flux is not reported by \cite{Miotello2021}.}
We derive the radius $R_{\text{CO}, 90\%}$  from the elliptical Gaussian approximation $R_{90\%} \simeq 1.42\  R_{68\%}$ (see \citealt{Sanchis2021}).

\subsection{Complete dataset}
\begin{figure*}[ht!]
    \centering
    \includegraphics[width=0.9\linewidth]{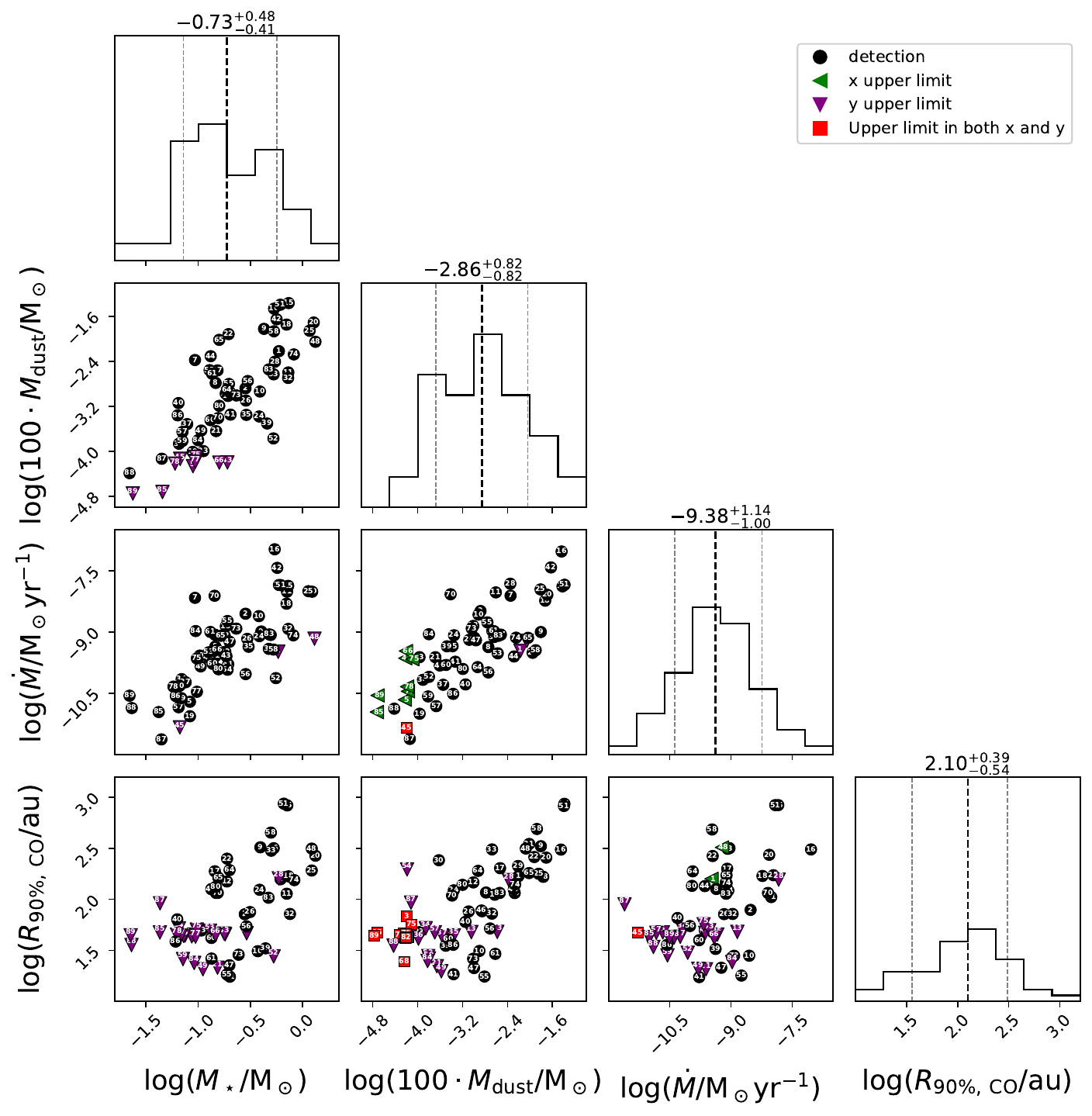} 
     
    \caption{\small{Distribution of the Lupus demographics across the four observables stellar mass $\log (M_\star / \text{M}_\odot)$, disc mass, obtained as 100 times the dust mass $\log (100 \cdot M_\text{dust} / \text{M}_\odot)$, mass accretion rate $\log (\dot{M} / \text{M}_\odot \ \text{yr}^{-1})$, and disc radius $\log (R_{\text{CO}, 90\%} / \text{au})$. Histograms represent the 1-dimensional distribution, where the median and the difference with respect to the $16^{\text{th}}$ and $84^\text{th}$ percentile are indicated. \rev{Percentiles are indicated by the dashed vertical lines}. The scatter plots show the correlation among the observables. Black circles represent detections, green and purple triangles the upper limits with respect to the x and y axes, respectively, and the red squares upper limits for both axes. Each point is labeled with the characteristic index of each source, defined in Table \ref{tab:Lupus}.}}
    \label{fig:Lupus_demographics}
\end{figure*}

We consider as dataset the distribution of the Lupus demographics in the four-dimensional observable space, defined by $\log (M_\star / \text{M}_\odot)$, $\log (M / \text{M}_\odot)$, $\log (\dot{M} / \text{M}_\odot \ \text{yr}^{-1})$, and $\log (R_{\text{CO}, 90\%} / \text{au})$.
Table \ref{tab:Lupus}, in the Appendix, lists the sources considered in this work together with their respective values in the four observables. 
Figure \ref{fig:Lupus_demographics} shows the distribution of the observed objects across the four-dimensional space. The numbers above the histograms indicate the median and the difference with respect to the $16^{\text{th}}$ and $84^\text{th}$ percentile without considering upper limits. 
Each object in the scatter plot is labeled by its corresponding index, defined in table \ref{tab:Lupus}.
Black circles represent detections, while green triangles, purple triangles and red squares represent upper limits for x, y and both axes, respectively.
The plot shows that the population exhibits a positive correlation among all four observables, although with a significant scatter (see also \citealt{Manara2023}).
Furthermore, we take into account the fraction of discs in the Lupus star-forming region.
We estimate this fraction by considering the number of members outlined by the \textit{Gaia} survey of \cite{Luhman2020} of 191, which reduces to 169 by excluding the confirmed binary systems.
We divide the number of sources in our disc sample by this number, obtaining a fraction of $f = 89/169$.
This fraction is still consistent with the one obtained by the more recent survey \cite{Polnitzky2026}.
We aim to exploit the full richness of the observed demographics, illustrated in Figure \ref{fig:Lupus_demographics}, by determining which model best reproduces their overall statistical properties.

\subsection{Adopted uncertainties}
\label{uncertainties}

We assumed an uncertainty of 0.1 dex for all measurements of $\log (M_\star / \text{M}_\odot)$, which is the typical error associated with PMS tracks models.
This uncertainty is also adopted for measurements of $\dot{M}$ (e.g., \citealt{Alcala2017}).
The main sources for uncertainty on the accretion rates come from uncertainty in stellar parameters and accretion variability: a combination of these contributions gives an uncertainty of 0.35 dex on $\log (\dot{M}/\text{M}_\odot \ \text{yr}^{-1})$ \citep{Alcala2014, Alcala2017, Manara2023}.
In contrast, assigning an uncertainty to the disc mass is not trivial, because it is not dominated by flux measurement errors, but rather by systematic uncertainties due to assumptions on the optical depth, the opacity and the gas-to-dust ratio.
A similar argument is valid for disc radius estimates, which strongly rely on CO abundance.
To set a realistic uncertainty for $M$, we compute the standard deviation between the dynamical measurements of $M$ by \cite{Longarini2025}, obtained within the exoALMA large program \citep{Teague2025}, and the dust-based estimates for the same sources. 
This yields $\sigma_{\log{M}} = 0.5$ dex.
We apply the same procedure for the $R_{90\% \text{CO}}$ and obtain $\sigma_{\log{R_{90\%,\text{CO}}}} = 0.25$ dex.
We adopt these uncertainties for all sources.

\section{Method and disc evolution models}
\label{model}

We apply our tool \texttt{DEMOS}, described in Paper I, to the Lupus sample, as described in Section \ref{observations} aiming to find the model parameters that most accurately reproduce the observed demographics for both the viscous-photoevaporative framework \citep{LyndenBellPringle1974, Clarke2001} and the MHD winds model \citep{BlandfordPayne1982, Tabone2022model}, and perform a statistically robust model comparison.
For each set of model parameters $\boldsymbol{\theta}$, we generate a synthetic population of 500 discs with the 1D disc population synthesis code \texttt{Diskpop} \citep{Somigliana2024}; (see Section 2.1 of Paper I for a detailed description).
$\boldsymbol{\theta}$ includes both parameters defining initial conditions and parameters defining the evolutionary model.
We sample the stellar masses $M_\star$ of synthetic objects from the Lupus stellar mass function, constructed by combining the $M_\star$ distribution of disc-bearing sources in the PPVII sample \citep{Manara2023} with that of disc-less sources. For the latter, spectral types are available for a subset of the sample in \cite{Luhman2020} and are converted to stellar masses following \cite{Baraffe2015}.
For each disc, initial conditions are extracted with Monte-Carlo methods, following distributions regulated by $\boldsymbol{\theta}$. 
Then, the disc surface density $\Sigma$ is evolved according to the selected evolutionary model until an age $t$. 
We consider synthetic discs dispersed, and remove them from the population, if either $M$ or $\dot{M}$ drops below the observational threshold, which we set to $10^{-6}$ M$_\odot$ for the disc mass and $10^{-12}$ M$_\odot$ yr$^{-1}$ for the accretion rate \citep{Manara2013_threshold}.
Subsequently, we retrieve global disc properties from the $\Sigma$ profile of the survived discs.
To obtain the radius that encloses 90\% of the CO flux from synthetic discs, we use the analytical formula derived by \cite{Trapman2023} 
\begin{equation}
    N_\text{H}(R_\text{90\%CO}) = 3.7 \cdot 10^{21} \ \delta_\text{C}^{-1} \left(M/{\text{M}_\odot}\right)^{0.34} \ \text{cm}^{-2} \ ,
\end{equation}
where $\delta_\text{C}$ is the carbon depletion factor \citep{Toci2023}, which we set to 0.2, following \cite{Trapman2025_AGE_PRO} and \cite{Tabone2025}.
The column density $N_\text{H}$ can be obtained using $N_\text{H} = \Sigma / \mu_\text{gas}$, where $\mu_\text{gas}$ is the mean molecular weight, commonly set in protoplanetary discs to 2.3 times the mass of an hydrogen atom.
Hence, we obtain a synthetic population $S$ whose distribution in the four-dimensional observables space approximates the expectations from the model $M$ and the parameters $\boldsymbol{\theta}$.
We evaluate the likelihood $p(D|\boldsymbol{\theta}, M)$, where $D$ is the dataset, following the procedure described in Section 2.2 of Paper I, taking into account also the Lupus disc fraction.
Then, we obtain the posterior distribution of $\boldsymbol{\theta}$ with  and \texttt{emcee} \citep{Foreman-Mackey2013} and the Bayesian evidence $p(D|M)$, which is required for model comparison, with the nested sampling tool \texttt{UltraNest} \citep{Buchner2021}. We additionally employ \texttt{UltraNest} to validate the posterior distributions obtained with \texttt{emcee}.

\subsection{The viscous-photoevaporative model}
\label{viscous_model}
In the turbulence-driven scenario, angular momentum is transported throughout the disc by turbulence, whose efficiency is parameterised by the $\alpha$-prescription by \cite{ShakuraSunyaev1973}, often assumed  constant with time and throughout the disc.
The evolution of $\Sigma$ in viscously accreting discs follows the self-similar solution by \cite{LyndenBellPringle1974}, which shows that the disc radius is expected to increase with time, due to the redistribution of angular momentum.
A purely viscous evolution does not produce the observed rapid disc dispersal \citep{Hernandez2007, Fedele2010}, a phenomenon which is commonly attributed to internal photoevaporation \citep{Hollenbach1994, Alexander2014}. 
\cite{Clarke2001} has shown that internal photoevaporation can carve cavities in discs, causing a rapid dispersal of the inner disc.
The overall evolution of the surface density driven by viscous accretion and photoevaporative dispersal can be expressed with an analytical solution, obtained by \cite{Ruden2004},
\begin{align}
    \Sigma (r,t) = \frac{M_0}{2 \pi R_0^2}\left(\frac{r}{R_0}\right)^{-\gamma} \left(1+\frac{t}{t_\nu}\right)^{-\frac{5/2 - \gamma}{2-\gamma}} \exp\left(-\frac{(r/R_0)^{{2-\gamma}}}{1+t/t_\nu}\right) \notag \\
    - \Delta \Sigma_\text{w}\left(r,t, \dot{\Sigma}_\text{w}  (\dot{M}_\text{w}, M_\star, r), t_\nu \right ) \ ,
\end{align}
where $\gamma<2$ is the scaling of the correlation between the kinematic viscosity $\nu = \alpha r (H/r)^2 \sqrt{GM_\star/r}$ and $r$, $R_0$ is the initial truncation radius, $M_0$ the initial disc mass, $t_\nu$ the typical viscous timescale, defined by $t_\nu = R_0^2/3(2-\gamma)\nu(R_0)$, and $\Delta \Sigma_\text{w}$ the surface depletion term due to internal photoevaporation (see Section 3 of Paper I).
$\Delta \Sigma_\text{w}$ is a function of $t$, $t_\nu$ and the wind's mass-loss rate profile $\dot{\Sigma}_\text{w}$, which further depends on the stellar mass $M_\star$, the total wind's mass-loss rate $\dot{M}_\text{w}$ and the radial location $r$.
The dispersal timescale can be obtained as a function of the disc initial properties, $M_\star$ and $\dot{M}_\text{w}$ \citep{Malanga2025}.
The effect of viscous spreading is encapsulated in the time dependence of the exponential cutoff, as the truncation radius expands with time.
We set $\gamma = 1$ and the aspect ratio
\begin{equation}
   H/r= 1/30 \ \left(M_\star/\text{M}_\odot\right)^{-1/2} \ \left(r/\text{au}\right)^{1/4}\ ,
    \label{eqn:hr}
\end{equation}
which follows the temperature $T \propto r^{-1/2}$ prescription \citep{Hartmann1998}, expected for irradiated discs.
As for internal photoevaporation, we adopt the mass-loss profile derived by \cite{Owen2012} via hydrodynamical simulations.

\subsection{The MHD winds model}

According to the MHD winds model, magnetic winds launched from the disc surface apply a torque to the disc, causing both accretion and dispersal.
\cite{Tabone2022model} obtained an analytical solution \rev{to describe} the evolution of discs experiencing this process, which is based on the generalisation of the $\alpha$-parameter \citep{ShakuraSunyaev1973}, namely $\alpha_\text{DW}$ and the magnetic lever arm $\lambda$.
We consider the pure wind solution where $\alpha$ depends on time following the relation
\begin{equation}
    \alpha_\text{DW}(t) = \alpha_\text{DW}(0) \left(\Sigma_\text{c}(t)/\Sigma_\text{c}(0)\right)^{-\omega} \ , 
\end{equation}
where $\Sigma_\text{c}$ is the surface density at the truncation radius and $\omega$, between 0 and 1, quantifies the dissipation of the magnetic field.
According to this solution, the surface density follows
\begin{equation}
    \Sigma(r,t) = \frac{M_0}{2 \pi R_0^2} \left(1-\frac{\omega}{2 t_\text{acc,0}} t \right)^{1/\omega} \left(\frac{r}{R_0}\right)^{-1+\xi} \exp(-r/R_0) \,
\end{equation}
where $\xi \equiv 1/[2(\lambda-1)]$ and $t_\text{acc,0}$ is the typical accretion timescale, defined analogously to the viscous case with $\alpha_\text{DW}$ replacing $\alpha$.
This solution leads to the full dispersal of the disc in a finite time $t_\text{disp} = 2 t_\text{acc,0}/\omega$, as long as $\omega \neq 0$. 
Like in the viscous model, for the MHD winds model, we adopt the prescription for the aspect ratio $H/r$ defined in equation \eqref{eqn:hr}.
The exponent $\xi$ can also be expressed as a function of the ejection-to-accretion ratio $f_M \equiv \dot{M}_\text{mhd,w}/\dot{M}$ with the following relation,
\begin{equation}
    f_M + 1 = \left(R_0/R_\text{in}\right)^\xi \,
\end{equation}
where $R_\text{in}$ is the disc inner radius, set to 0.03 au, and $\dot{M}_\text{mhd,w}$ is the mass-loss rate due to MHD windss.

\subsection{Free parameters}
For each model, we aim to obtain the posterior distribution of the characteristic parameters. 
We assume that initial disc properties, such as initial mass $M_0$ and initial size $R_0$, follow power-law correlations with the stellar mass.
Under this assumption, a generic disc property $q_i$ is described by
\begin{equation}
    q_i = q' \cdot  \left(\frac{M_{\star,i}}{0.3 \ \text{M}_\odot}\right)^{\lambda_q} \cdot 10^{\mathcal{N}(0,\sigma_q)} \ ,
\end{equation}
\citep{Somigliana2022} where $\mathcal{N}(0,\sigma_q)$ identifies a normal distribution with spread $\sigma_q$, $q'$ is the median value of the property for $M_\star = 0.3 $ M$_\odot$ and $\lambda_q$ the slope of the correlation.
Similarly, we assume that the angular momentum efficiency parameters, namely $\alpha$ for the viscous framework and $\alpha_\text{DW}$ for the wind-driven model, also follow power-law correlations with $M_\star$.
For the viscous-photoevaporative framework, we further assume that the total mass-loss rate $\dot{M}_\text{w}$ follows a power-law correlation with $M_\star$.
For all these quantities, we treat the normalisation (value for $M_\star$ =  $0.3\ \text{M}_\odot$), the slope and the spread as free parameters.
In addition, for the MHD winds model, we also include $\omega$ and $f_M$ in the set of free parameters.
They, however, do not correlate with $M_\star$; but have the same value for all discs in the population.
Lastly, we consider the mean age $\langle t\rangle$ and the intrinsic age scatter $\sigma_t$ as free parameters.
Table \ref{tab:free_parameters} summarises the adopted free parameters for both models, where for quantities correlated with $M_\star$ we constrain normalisation, slope and spread.
\begin{table}[t]
\centering
\scriptsize
\renewcommand{\arraystretch}{1.1}

\begin{tabularx}{\columnwidth}{l c X c c}
\toprule
\textbf{Quantity} & \textbf{Correlated with $M_\star$?} & \textbf{Parameter} & \textbf{Symbol} & Prior \\
\midrule

\rowcolor{gray!15}
\multicolumn{5}{l}{\textbf{Both models}} \\

$M_0$ [M$_\odot$] & Yes & Normalisation & $\log M_0'$ & $\mathcal{U}[-5,0]$ \\
 & & Slope & $\lambda_{M_0}$ & $\mathcal{U}[-4,4]$ \\
 & & Spread & $\sigma_{M_0}$ & $\mathcal{U}[0,1]$ \\

\rowcolor{gray!5}
$R_0$ [au] & Yes & Normalisation & $\log R_0'$ & $\mathcal{U}[0,2]$ \\
\rowcolor{gray!5}
 & & Slope & $\lambda_{R_0}$ & $\mathcal{U}[-4,4]$ \\
\rowcolor{gray!5}
 & & Spread & $\sigma_{R_0}$ & $\mathcal{U}[0,1]$ \\

$t$ [Myr] & No & Mean & $\langle t \rangle$ & $\mathcal{N}[2,0.6]$ \\
 & & Scatter & $\sigma_t$ &  $\mathcal{U}[0,0.7]$ \\

\midrule
\rowcolor{blue!15}
\multicolumn{5}{l}{\textbf{Viscous–photoevaporative model}} \\

$\alpha$ & Yes & Normalisation & $\log\alpha'$ & $\mathcal{U}[-5,-1]$ \\
 & & Slope & $\lambda_{\alpha}$ & $\mathcal{U}[-4,4]$ \\
 & & Spread & $\sigma_{\alpha}$ & $\mathcal{U}[0,1]$ \\

\rowcolor{blue!5}
$\dot{M}_\text{w}$ [M$_\odot$ yr$^{-1}$] & Yes & Normalisation & $\log \dot{M}_\text{w}'$ & $\mathcal{U}[-11,-7]$ \\
\rowcolor{blue!5}
 & & Slope & $\lambda_\text{w}$ & $\mathcal{U}[-4,4]$ \\
\rowcolor{blue!5}
 & & Spread & $\sigma_\text{w}$ & $\mathcal{U}[0,1]$ \\

\midrule
\rowcolor{orange!15}
\multicolumn{5}{l}{\textbf{MHD winds model}} \\

$\alpha_\text{DW}$ & Yes & Normalisation & $\log \alpha_\text{DW}'$ & $\mathcal{U}[-5,-1]$ \\
 & & Slope & $\lambda_{\alpha_\text{DW}}$ & $\mathcal{U}[-4,4]$ \\
 & & Spread & $\sigma_{\alpha_\text{DW}}$ & $\mathcal{U}[0,1]$ \\

\rowcolor{orange!5}
$\omega$ & No & Fixed value & $\omega$ & $\mathcal{U}[0,1]$ \\

$f_M$ & No & Fixed value & $\log f_M$ & $\mathcal{U}[-1,1]$ \\

\bottomrule
\end{tabularx}

\caption{\small{Free parameters of the viscous-photoevaporative and MHD winds models. Quantities correlated with $M_\star$ (e.g., $M_0$) include slopes, normalisations, and spreads as free parameters. Quantities kept constant across the population (e.g., $\omega$) are treated as fixed. $\mathcal{U}[l,u]$ denotes a uniform prior with lower limit $l$ and upper limit $u$, while $\mathcal{N}[\mu,s]$ denotes a normal prior with mean $\mu$ and spread $s$. In total, 14 parameters are considered for the viscous-photoevaporative model and 13 for the MHD winds model.}}
\label{tab:free_parameters}

\end{table}
We further highlight the adopted prior distribution for all parameters. 
We set uniform prior for all parameters, except for the mean age $\langle t \rangle$, where independent estimates in the literature indicate $t \sim 1$–$3$ Myr \citep{Alcala2017, Galli2020_Lupus} \rev{and 2 Myr is typically considered in population synthesis studies (e.g., \citealt{Tabone2022, Emsenhuber2023, Tabone2025})}.
The number of free parameters amounts to 14 for the viscous-photoevaporative model and to 13 for the MHD winds model.

\section{Results}
\label{results}
\begin{figure}[h!]
    \centering
    
    \begin{subfigure}{0.96\hsize}
        \includegraphics[width=\hsize]{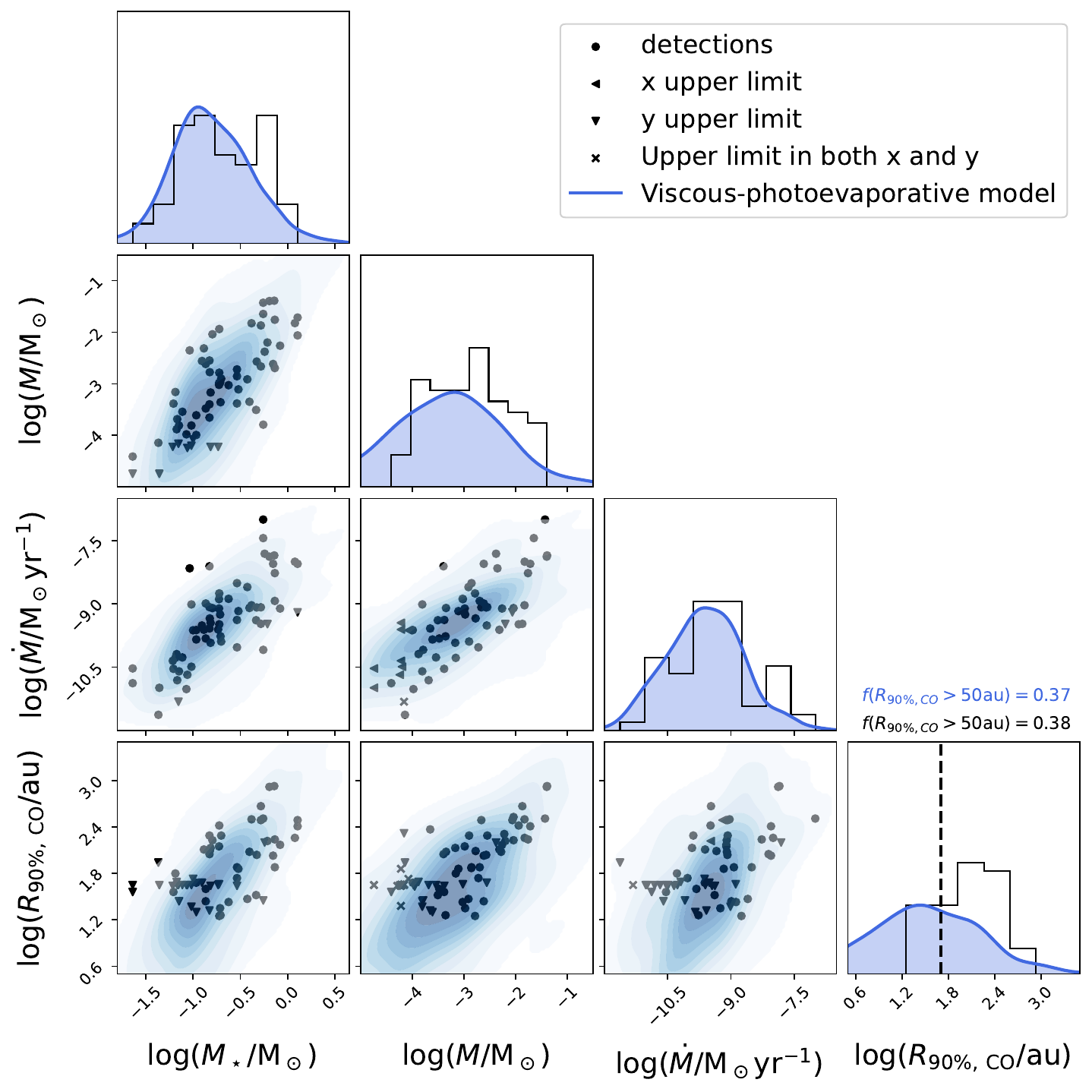}
        \caption{}
    \end{subfigure}
    
    \vspace{0.3cm}
    
    \begin{subfigure}{0.96\hsize}
        \includegraphics[width=\hsize]{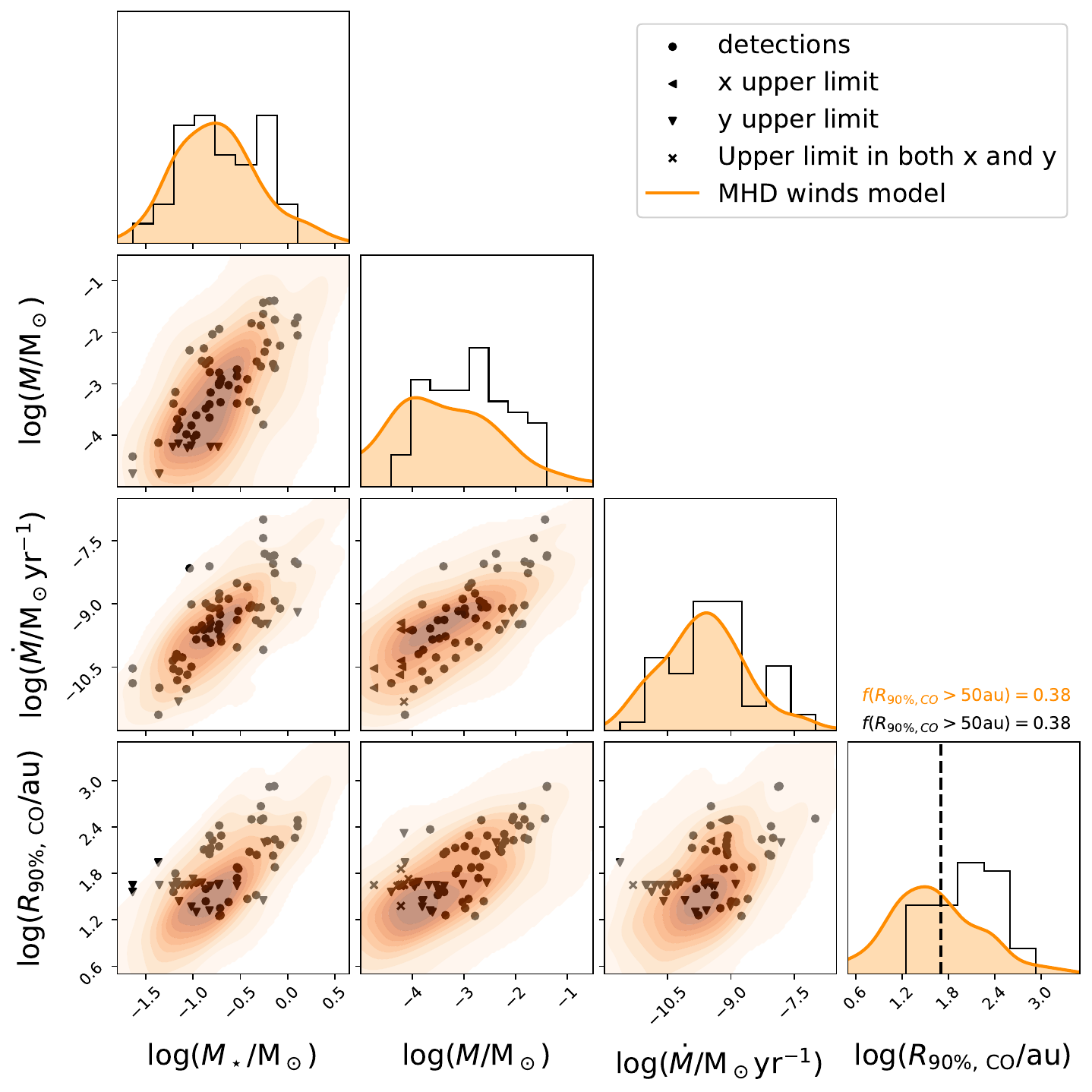}
        \caption{}
    \end{subfigure}
    
    \caption{\small{
    Comparison between the observed Lupus distribution in the observables space (black), where the axes are $\log (M_\star/$M$_\odot)$, $\log (M/$M$_\odot)$, $\log (\dot{M}/$M$_\odot $yr$^{-1})$ and $\log (R_{90\%, \text{CO}}/$au) to the probability function predicted by the viscous model (blue, top panel) and MHD winds model (orange, bottom panel). Dots represent detections, while triangles and crosses identify measurements with an upper limit on one or both axes, respectively.
    The dashed vertical lines in the $\log (R_{90\%, \text{CO}}/$au) histogram indicate the typical detection threshold. The fraction of discs with size above such threshold is reported above the histogram for both observed discs (black) and best-fit synthetic discs (blue and orange).
    }}
    \label{fig:fit}
\end{figure}

We manage to reproduce the Lupus demographics remarkably well with both the viscous-photoevaporative and the MHD winds models.
Figure \ref{fig:fit} shows a comparison between the Lupus data (black points) and the probability curves predicted by the models.
We obtained the theoretical probability distributions of the viscous (blue curves, top panel) and wind-driven (orange curves, bottom panel) models by computing the Kernel Distribution Estimation (KDE) \citep{Rosenblatt1956, Parzen1962} of the best-fit synthetic populations, where a scatter is added to synthetic points to mimic observational uncertainties.
The histograms on the diagonal show the comparison between the 1D distribution of observed quantities, excluding upper limits, with the respective model predictions.
The figure shows that both evolutionary frameworks reproduce the observed demographics in the observables space.
The size distribution predicted by the model differs from the observed one due to the significant impact of upper limits.
However, the fraction of synthetic discs with size above the typical detection threshold is compatible with the one of observed discs.
We constrain the model parameters that best reproduce the observed demographics with significant precision.
Table \ref{tab:results} shows the posterior distribution $p(\boldsymbol{\theta}|D, M)$ of the model parameters for both the viscous-photoevaporative and the MHD winds frameworks.
The table shows the median value with uncertainties corresponding to the 16$^\text{th}$ and 84$^\text{th}$ percentile of the distribution.
We show the whole distributions in Appendix \ref{Appendix_results}.
\rev{
The corner plots in Appendix \ref{Appendix_results} show that some parameters of the model (in particular the spreads of the initial distributions) are poorly constrained by the data. 
We thus attempted a fit by fixing them to their median posterior value to see if that improved the quality of the fit. 
As shown in Appendix \ref{test_spread}, this is not the case, as both the fit quality and the uncertainties on the remaining parameters are left unchanged.
In Appendix \ref{test_spread}, we also discuss the consequences or choosing different priors.}
\begin{table}[h!]
\centering
\scriptsize
\renewcommand{\arraystretch}{1.35}
\setlength{\extrarowheight}{1.5pt}

\begin{tabularx}{0.5\textwidth}{l X l X}

\toprule
\multicolumn{4}{c}{\textbf{Results}} \\
\midrule

\multicolumn{2}{c}{\cellcolor{blue!15}\textbf{Viscous-photoevaporative model}} &
\multicolumn{2}{c}{\cellcolor{orange!15}\textbf{MHD winds model}} \\

\cmidrule(lr){1-2} \cmidrule(lr){3-4}

\cellcolor{blue!5}\textbf{Parameter} & 
\cellcolor{blue!5}\textbf{16$^{\mathrm{th}}$, 50$^{\mathrm{th}}$, 84$^{\mathrm{th}}$} &
\cellcolor{orange!5}\textbf{Parameter} & 
\cellcolor{orange!5}\textbf{16$^{\mathrm{th}}$, 50$^{\mathrm{th}}$, 84$^{\mathrm{th}}$} \\

\midrule

$\log M_0'$ & $-2.3^{+0.2}_{-0.2}$ & $\log M_0'$ & $-2.3^{+0.2}_{-0.2}$ \\
$\lambda_{M_0}$ & $1.8^{+0.3}_{-0.4}$ & $\lambda_{M_0}$ & $2.0^{+0.2}_{-0.2}$ \\
$\sigma_{M_0}$ & $0.3^{+0.2}_{-0.2}$ & $\sigma_{M_0}$ & $0.4^{+0.1}_{-0.1}$ \\
$\log R_0'$ & $0.4^{+0.3}_{-0.3}$ & $\log R_0'$ & $0.7^{+0.2}_{-0.2}$ \\
$\lambda_{R_0}$ & $2.2^{+0.6}_{-0.7}$ & $\lambda_{R_0}$ & $1.0^{+0.3}_{-0.3}$ \\
$\sigma_{R_0}$ & $0.6^{+0.2}_{-0.3}$ & $\sigma_{R_0}$ & $0.6^{+0.2}_{-0.1}$ \\
$\langle t\rangle$ [Myr] & $0.9^{+0.3}_{-0.3}$ & $\langle t\rangle$ [Myr] & $1.6^{+0.4}_{-0.4}$ \\
$\sigma_t$ & $0.5^{+0.2}_{-0.2}$ & $\sigma_t$ & $0.3^{+0.2}_{-0.2}$  \\
$\log \alpha'$ & $-3.5^{+0.2}_{-0.1}$ & $\log \alpha_\text{DW}'$ & $-3.4^{+0.2}_{-0.1}$ \\
$\lambda_\alpha$ & $2.0^{+0.6}_{-0.4}$ &  $\lambda_{\alpha_\text{DW}}$ & $2.0^{+0.3}_{-0.3}$\\
$\sigma_\alpha$ & $0.2^{+0.5}_{-0.1}$ &  $\sigma_{\alpha_\text{DW}}$ & $0.1^{+0.2}_{-0.1}$\\
$\log \dot{M}_w'$ & $-8.8^{+0.2}_{-0.3}$ & $\omega$ & $0.4^{+0.2}_{-0.2}$ \\
$\lambda_w$ & $2.6^{+0.5}_{-0.4}$ & $\log f_M$ & $-0.6^{+0.4}_{-0.3}$ \\
$\sigma_w$ & $0.4^{+0.4}_{-0.3}$ &  \\

\bottomrule
\end{tabularx}

\caption{\small{Summary of the inferred parameters for the viscous-photoevaporative and MHD winds models. Values are reported as the median with uncertainties corresponding to the 16$^\text{th}$ and 84$^\text{th}$ percentiles of the posterior distribution obtained with \texttt{DEMOS}.}}
\label{tab:results}

\end{table}

\subsection{Initial conditions}

\begin{figure*}[h!]
    \centering
    \includegraphics[width=0.7\linewidth]{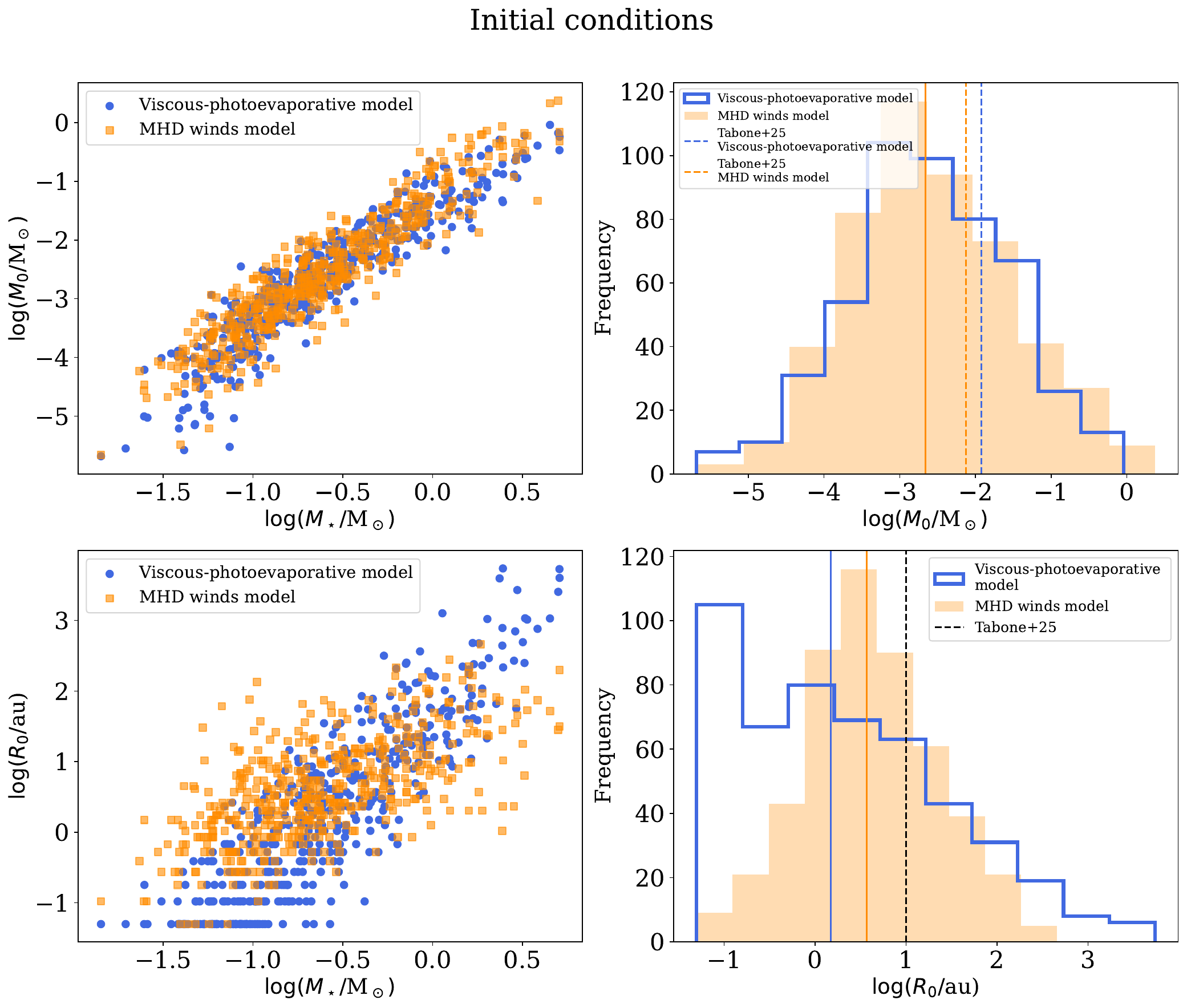} 
     
    \caption{Distributions of the initial mass and size of synthetic discs, generated using the parameters inferred by \texttt{DEMOS} from Lupus data. The top row shows the $M_0$ distribution, for both the viscous model (blue) and the MHD winds model (orange), whereas the bottom row shows the $R_0$ distribution. The left panels show the correlation between disc initial properties and the stellar mass, while the right panels show the 1D distributions, where the medians of synthetic populations (solid lines) are compared with the results from \citep{Tabone2025} (dashed lines).}
    \label{fig:initial_conditions}
\end{figure*}

The bulk of the population, corresponding to $M_\star = 0.3$ M$_\odot$, is characterised by initial disc masses $M_0 \sim 3-5 \times 10^{-3}$ M$_\odot$.
Moreover, $M_0$ exhibits a steep dependence on the stellar mass $M_\star$, following a power law relation with exponent 1.8 and 2.0, under the viscous and wind-driven scenarios, respectively.
These correlations are in agreement with the values estimated by \cite{Somigliana2022, Somigliana2024} and imply $M_0 \sim 10^{-2}-10^{-1}$  M$_\odot$ for discs surrounding 1 M$_\odot$ star, consistent with observations of massive discs \citep{Longarini2025}.
\texttt{DEMOS} also provides important constraints on the initial size of discs, showing that the bulk of the disc population is substantially compact, with initial truncation radius $R_0 \sim 3$ au and 5 au for the viscous and MHD winds models, respectively.
This result is sensitive to our assumption on $^{12}$CO non-detections, where, following \cite{Miotello2021}, we associated an upper limit in flux to an upper limit in radius.
The consequences of different assumptions on $R_{90\%, \text{CO}}$ non-detections are discussed in Section \ref{R_upper_limits}.
Our estimates of the initial disc sizes are not fully consistent with observations of Class 0 sources \citep{Maury2019}, which indicate that young protostellar discs are generally more extended. However, a direct comparison is complicated by the non-trivial relation between the observed Full Width at Half Maximum (FWHM) of the continuum emission and $R_0$ and the fact that Class 0 samples are biased towards bright sources.
Figure \ref{fig:initial_conditions} shows the initial conditions of two synthetic populations, generated with the parameters inferred by \texttt{DEMOS} for both the viscous model (blue) and the MHD winds model (orange).
The top row shows the initial mass distribution correlated with $M_\star$ (left) along with a comparison to \cite{Tabone2025} (right).
The higher median values obtained by \cite{Tabone2025} are due to the fact that the AGE-PRO sample is limited to stars with $M_\star \sim 0.5$ M$_\odot$, larger than the median $M_\star$ of our sample.
The bottom row shows the $R_0$ distribution. 
The bottom-left panel, illustrating the $R_0$–$M_\star$ correlation, highlights the steep scaling predicted by the viscous model. 
In contrast, the bottom-right panel shows that the viscous model is skewed toward smaller $R_0$ values, while still producing a more populated tail of extended discs.
The two models differ in their predicted initial correlation between discs sizes and $M_\star$; indeed, the MHD winds model suggests a linear scaling, whereas viscous evolution points to a roughly quadratic relation.
This stronger scaling and larger intrinsic spread cause the viscous model to predict initial disc sizes spanning nearly four orders of magnitude.
This discrepancy can be explained by the fact that, while physical disc sizes remain somewhat constant under pure wind-driven evolution \citep{Tabone2022model, Trapman2022_MHD}, the viscous model, coupled with internal photoevaporation, incorporates mechanisms that cause the disc to expand, such as viscous spreading, or shrink, such as outside-in dispersal (see \citealt{Malanga2025, Tabone2025}).
Due to the positive scalings with $M_\star$, discs around 1 M$_\odot$ stars are initially extended ($20 - 40$ au).
The marked difference in estimated initial disc sizes between solar-type stars and M dwarfs potentially has a strong impact on the relative ability of the two classes of stars in forming planets (see \citealt{Payne_2007} for a similar argument).
These initial conditions should be interpreted as properties that must already be present as soon as the disc is formed, and, therefore, need to be injected in the population during the protostellar phase.
This hypothesis, originally proposed by \cite{Alexander_Armitage2006} and \cite{Dullemond2006}, has now been tested thanks to \texttt{DEMOS}, which efficiently explores the parameter space while fully exploiting demographics.
Consequently, models that explore the early stages of planet formation should investigate whether such correlations are naturally reproduced.
In Appendix \ref{GR} we show that the initial disc population, generated with Lupus best-fit parameters, is expected not to be gravitationally unstable.

\subsection{Model parameters}
The simultaneous analysis of the $M$, $\dot{M}$ and $R$ distributions allow us to put constraints on the angular momentum transport parameters, namely $\alpha$ \citep{ShakuraSunyaev1973} for the viscous model and $\alpha_\text{DW}$ \citep{Tabone2022model} for the wind-driven model (see Section 4.2 of Paper I).
In the viscous scenario, \texttt{DEMOS} obtains $\alpha \sim 3 \times 10^{-4}$ for a typical disc surrounding a $0.3\ \text{M}_\odot$ star, which is consistent with empirical constraints on turbulence in the literature (see the review \citealt{Rosotti2023}).
\rev{The accretion timescale $t_\nu$, for these discs is 0.4 Myr, which means that the truncation radius increases by a factor 3 due to viscous spreading}.
In the wind-driven case, we find $\alpha_\text{DW} \sim 4 \times 10^{-4}$, which roughly corresponds to a disc magnetization $\beta_\text{mag} \sim 10^5$ (see e.g., \citealt{Armitage2013, Tabone2022model}).
In both cases, our estimates align with the results of \cite{Tabone2025}.
Moreover, we find that angular momentum transport is more efficient for discs surrounding massive stars ($\alpha \sim 10^{-3}- 10^{-2}$ for $1 \ \text{M}_\odot$ stars), with an almost quadratic scaling between $\alpha$ and $M_\star$ in both models.
As a consequence of this steep correlation, $\alpha$ spans nearly four orders of magnitude across the disc population.
The spreads of this correlation are compatible with 0 for both the viscous $\alpha$ and $\alpha_\text{DW}$, which means that a large spread in the $\alpha - M_\star$ correlation is not required to reproduce the observations.
The scaling $\lambda_{\alpha_\text{DW}} = 2$ for the MHD winds model implies that the accretion timescale $t_{\mathrm{acc},0}$ scales with $M_\star$ with an exponent $\lambda_{t_{\mathrm{acc},0}} = \lambda_{R_0} - \lambda_{\alpha_\text{DW}} + 1/2 = -0.5$. This negative correlation hints at an inverse scaling between $M_0$ and $t_{\mathrm{acc},0}$, consistent with the findings of \cite{Zallio2024}, who studied the $\dot{M}-M$ correlation in the MHD winds model.
The need for a positive $\alpha-M_\star$ scaling to reproduce the observed $\dot{M}-M_\star$ correlation was first highlighted by \cite{Gorti2009b} and recently revived by \cite{Gomez2025}.
The initial slope of the $\dot{M}-M_\star$ correlation can be derived as $\lambda_{\dot{M}_0} = \lambda_{M_0} - \lambda_{R_0} + \lambda_\alpha - 1/2$, following the relation
\begin{equation}
    \dot{M}_0 = \frac{M_0}{2 t_\text{acc,0}} =\propto\frac{M_0}{R_0 \ M_\star^{1/2}}\alpha \ .
\end{equation}
By substituting our best-fit results, we obtain $\lambda_{\dot{M}_0} \sim 2.5$ for the MHD winds model, which aligns with the assumption of \cite{Alexander_Armitage2006} and \cite{Dullemond2006} that the observed $\dot{M} \propto M_\star^2$ correlation arises from initial conditions.
On the other hand, in the viscous model $\lambda_{\dot{M}_0} \sim 1.1$ suggesting that the $\dot{M} - M_\star$ slope steepens with time due to viscous evolution, as predicted by \cite{Somigliana2022} (see Appendix \ref{slopes} for a detailed discussion) \rev{and that discs surrounding massive stars spread more slowly}.

With \texttt{DEMOS}, it is also possible to constrain indirectly the wind's mass-loss rate $\dot{M}_\text{w}$, whose intensity sets the disc dispersal time in the viscous model \citep{Clarke2001, Picogna2021, Malanga2025}.
Indeed, the inclusion of the disc fraction in Lupus among the observables plays a key role in constraining this parameter.
We find that the bulk of the population experiences $\dot{M}_\text{w} \sim 1.4 \times 10^{-9}$ M$_\odot$ yr$^{-1}$, in agreement with estimates from state-of-the-art hydrodynamical simulations \citep{Nakatani2018, Sellek2024b}.
Including the $M_\star$ of Lupus discs among the observables provides important information about how $\dot{M}_\text{w}$ depends on $M_\star$.
This is because the $M_\star$ distribution of all synthetic populations is sampled from the Lupus stellar mass function, including both disc-bearing and disc-less stars.
Since $M_\star$ does not evolve in our simulations, \texttt{DEMOS} matches the observed stellar-mass distribution by adjusting the slope of the $\dot{M}_\text{w}$–$M_\star$ relation, $\lambda_\text{w}$, which controls the disc fraction as a function of $M_\star$.
As a consequence, our results reproduce the observed trend that discs around massive stars are more short-lived \citep{Carpenter2006, Yasui2014, Ribas2015, Richert2018, Pfalzner2022}.
We find that the data are best reproduced by $\lambda_\text{w} \sim 2.6$, a much steeper scaling than what predicted by hydrodynamical simulations \citep{Picogna2021}.
However, the intrinsic scatter of this correlation is significant.
In Appendix \ref{disc_fraction_stellar_mass} we show the comparison between the synthetic and observed disc fraction as a function of $M_\star$.

For what concerns the MHD winds model, we further obtain a low ejection-to-accretion ratio of $\sim 0.4$ and $\omega \sim 0.4$.

\subsection{Estimated ages}
\label{ages}

Perhaps the most significant result of this inference concerns the estimated mean age of Lupus, which we decided to include among the free parameters due to the loose observational constraints from stellar evolutionary tracks.
Indeed, while the MHD winds-driven evolution points to $\langle t \rangle = 1.6^{+0.4}_{-0.4}$ Myr, compatible within 1$\sigma$ to the typically assumed age of 2 Myr \citep{Alcala2017, Galli2020_Lupus}, the viscous model instead yields $\langle t \rangle = 0.9^{+0.3}_{-0.3}$ Myr, significantly lower than standard estimates.
This occurs because \texttt{DEMOS} finds a better agreement with observed demographics when considering younger synthetic population evolved under the viscous model, despite the prior distribution of $\langle t \rangle$ being centred at 2 Myr.
This result is not surprising, as it reflects an intrinsic property of turbulent-driven accretion.
Indeed, in the purely viscous solution \citep{LyndenBellPringle1974} the ratio between disc mass and accretion rate can be expressed as 
\begin{equation}
    \frac{M}{\dot{M}} = 2 (2-\gamma)(t+t_\nu) \gtrsim 2t \ ,
    \label{eqn:viscous_timescale}
\end{equation}
implying that, in the $\gamma = 1$ case, $M/\dot{M}$ is always lower than twice the disc age and asymptotically approaches this value in the limit $t \gg t_\nu$ \citep{Jones2012, Rosotti2017}.
However, in the Lupus population, more than 60\% of the sources have $M/\dot{M}< 4$ Myr, showing that viscous evolution does not reproduce Lupus demographics if we set $t = 2$ Myr.
Such conclusion was the gist of \cite{Tabone2025}, who found a poor agreement between the viscous model and the AGE-PRO data from Lupus, having fixed the age to 2 Myr.
On the other hand, our result is in significant agreement with \cite{Lodato2017}, who obtained that ages $t \lesssim 1$ Myr provide a better reproduction of the Lupus $\dot{M}- M$ correlation, although their available dataset was a subset of the one that we considered.
In contrast, our estimated ages are not compatible with findings of recent \textit{Gaia} surveys \citep{Polnitzky2026}, who obtain $\langle t\rangle \sim 6$ Myr.

\begin{figure*}[h!]
    \centering

    \begin{subfigure}{0.4\textwidth}
        \centering
        \includegraphics[width=\hsize]{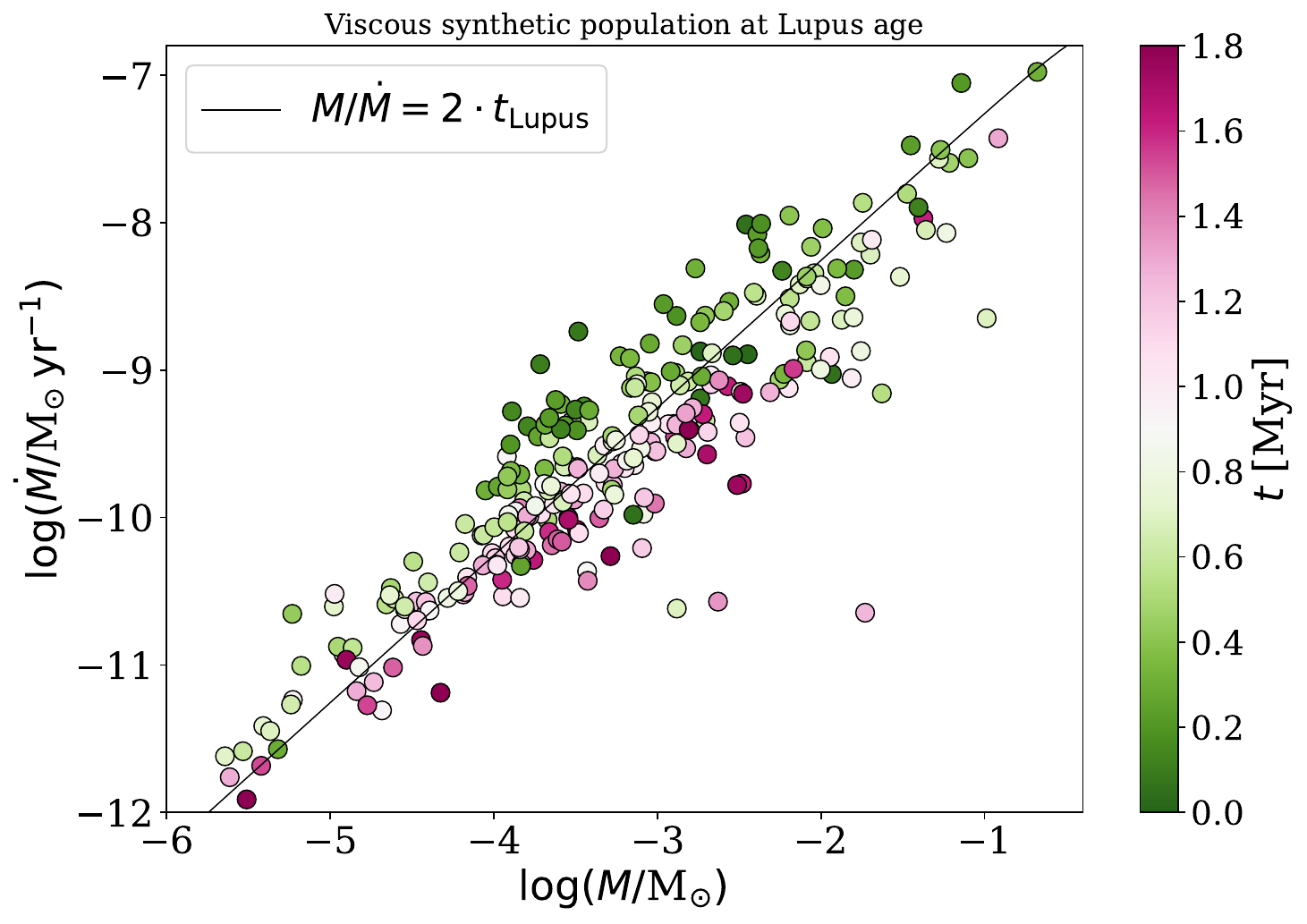}
        \caption{}
    \end{subfigure}
    \begin{subfigure}{0.4\textwidth}
        \centering
        \includegraphics[width=\hsize]{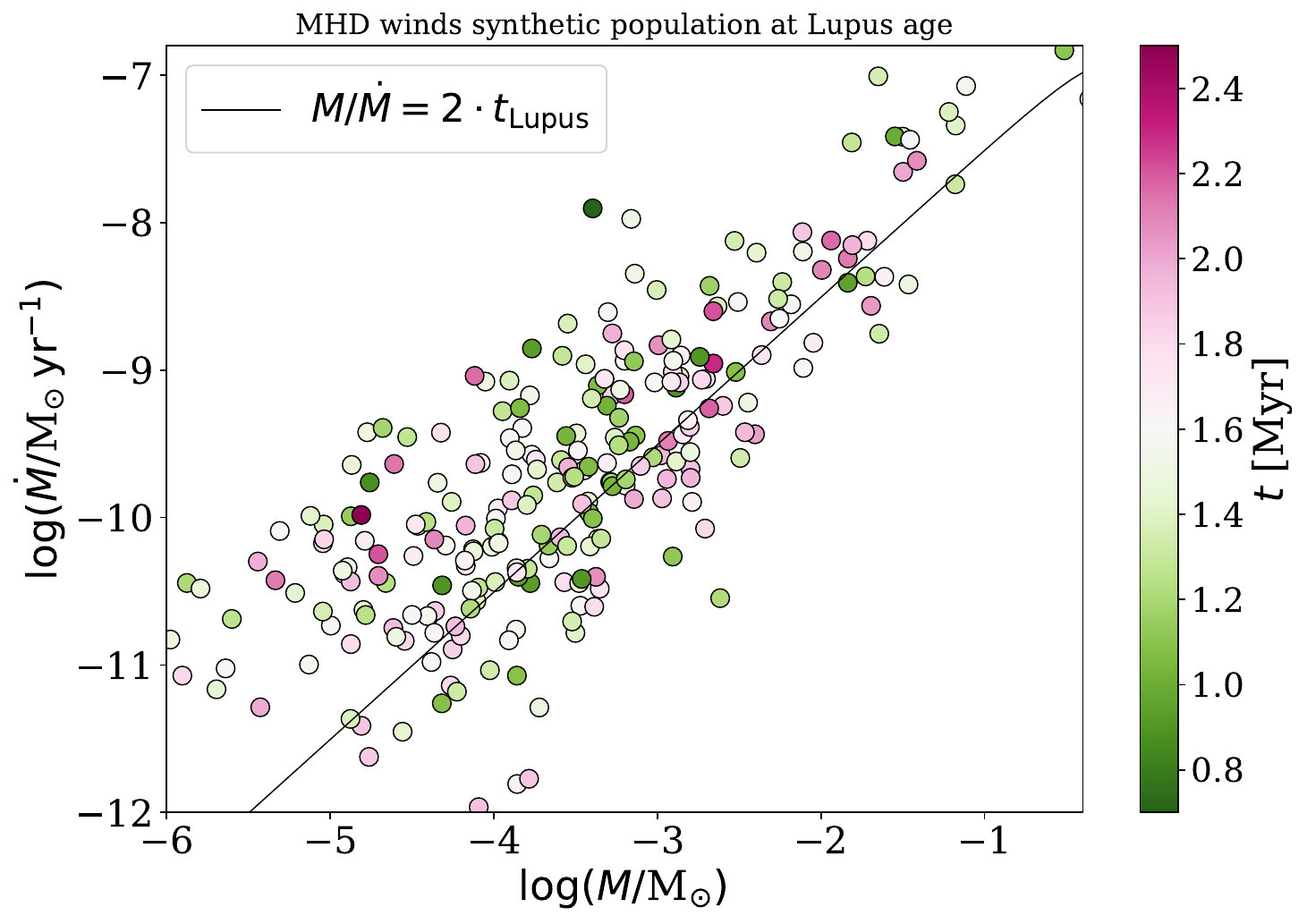}
        \caption{}
    \end{subfigure}

    \caption{\small{
    Synthetic populations evolved with the best-fit parameters obtained by \texttt{DEMOS} for the viscous model (left) and the MHD winds model (right), respectively, displayed in the $\dot{M}-M$ plane. The colorbar highlights the age of individual discs, where green and pink correspond to discs younger and older than the mean age, respectively. The mean ages of the population are 0.9 Myr in the viscous case and 1.6 Myr in the wind-driven case.
    The black line traces the viscous isochrone.
    }}
    \label{fig:age_scatter}
\end{figure*}

In addition, the two models show a discrepancy in the estimates of the age scatter $\sigma_t$.
The viscous model shows a preference for $\sigma_t \sim 0.5$ Myr, while in the MHD winds inference this parameter remains unconstrained (see also Appendix \ref{Appendix_results}).
This is due to the fact that the viscous model predicts a lower spread in $\dot{M}/M$ \citep{Lodato2017, Mulders2017, Somigliana2023} with respect to the MHD winds framework.
Consequently, to reproduce the large observed spread, the viscous model strongly relies on the age scatter.
Figure \ref{fig:age_scatter} shows two synthetic populations obtained with the best-fit model parameters of the viscous \rev{(a)} and MHD winds models \rev{(b)}, respectively.
Each disc is evolved until an age $t$ extracted from a normal distribution with mean $\langle t \rangle$ and spread $\sigma_t$, where the values differ between the two evolutionary scenarios.
Figure \ref{fig:age_scatter} shows the $\dot{M}-M$ plane, compared to the viscous isochrone $M/\dot{M} = 2 \langle t \rangle$. The colorbar highlights the age of individual discs, where green and pink dots correspond to discs younger and older than the mean age, respectively.
The plots highlight that in the viscous case, green dots generally lie above the viscous isochrone, while pink ones are placed below it.
On the other hand, the MHD winds model does not manifest any age gradient.
This emphasises the fact that the viscous model requires an age scatter to explain the observed spread in $M/\dot{M}$.
Moreover, the plots show that the age of single YSOs could represent a precious source of information to disentangle the two evolutionary models.

\subsection{Overall model accuracy}
\label{model_accuracy}

One of the main advantages of our statistical framework is that it allows us to compute the Bayesian evidence $Z$, which quantifies how accurately models are reproducing the data. 
Indeed, the Jeffrey's scale \citep{Jeffreys1939} provides a metric to convert the ratio of evidences between two models into a relative probability, where a strong support to one model over the other is characterised by $|\ln B_{12}| = |\ln Z_1 -\ln Z_2| \gtrsim 5$, where $B_{12}$ is the Bayes factor.
We compute the evidence using \texttt{UltraNest} \citep{Buchner2021} and obtain $\ln Z_\text{viscous} = -255 \pm 5$ and $\ln Z_\text{MHD} = -259 \pm 3$, which indicates a slight preference for the viscous-photoevaporative framework.
However, due to the large uncertainties on the estimated evidence (see Appendix \ref{Appendix_results} for a detailed discussion) and to the low difference between $\ln Z_\text{viscous}$ and $\ln Z_\text{MHD}$, the Bayesian inference of Lupus demographics does not allow us to rule out either model.
Indeed, the accuracy with which the two evolutionary scenario explains the observed population is comparable (see Figure \ref{fig:fit}).
To further compare the goodness of the fit of the two models, we computed the Bayesian Information Criterion (BIC) \citep{Schwarz1978}, defined by
\begin{equation}
    \text{BIC} \equiv k \ln(n) - 2 \ln (\hat{L}) \ ,
\end{equation}
where $k$ is the number of free parameters, $n$ the number of data points and $\hat{L}$ the maximum likelihood.
We further validated the results by using the Deviance Information Criterion (DIC) \citep{Spiegelhalter2002}, defined by
\begin{equation}
    \text{DIC} \equiv -4 \ \overline{\ln (P)} + 2 \ln (\hat{P}) \ ,
\end{equation}
where $\overline{P}$ and $\hat{P}$ are the mean and maximum posteriors, respectively.
We obtain $\Delta \text{BIC} \equiv \text{BIC}_\text{viscous} - \text{BIC}_\text{MHD} \sim -9$ and  $\Delta \text{DIC} \sim -10$ where lower values indicate better fits, and differences larger than 6 are considered strong evidence in favour of one model over the other.
Both criteria show a strong preference for the viscous-photoevaporative model, highlighting that it reproduces the Lupus demographics with a higher accuracy. However, the low difference in Bayesian evidence does not allow us to rule out the MHD winds model.

Moreover, we performed a fit of the Lupus demographics with both models but keeping the mean age fixed at 2 Myr.
We obtain $\text{BIC}_\text{visous, 2 Myr} - \text{BIC}_\text{MHD, 2 Myr} \sim 9$ and $\text{DIC}_\text{visous, 2 Myr} - \text{DIC}_\text{MHD, 2 Myr} \sim 17$.
This result is consistent with the findings of \cite{Tabone2025}, that the viscous-photoevaporative model is less consistent with observed data if evaluated at 2 Myr.
This agreement is found despite the use of different datasets: \cite{Tabone2025} reproduced a smaller subset of Lupus \citep{Zhang2025, Deng2025} with disc masses inferred from CO and N$_2$H$^{+}$, whereas our $M$ data are derived from the continuum flux.
Moreover, the best-fit population obtained with the \texttt{DEMOS} fit with $\langle t \rangle = 2$ Myr shows substantial differences with the one obtained with flexible mean age.
These differences are described in Appendix \ref{viscous_model_2_Myr}.

\section{Discussion}
\label{discussion}

\subsection{Expected evolution with time}
\label{evolution}

\begin{figure*}[ht!]
    \centering
    \includegraphics[width=0.7\linewidth]{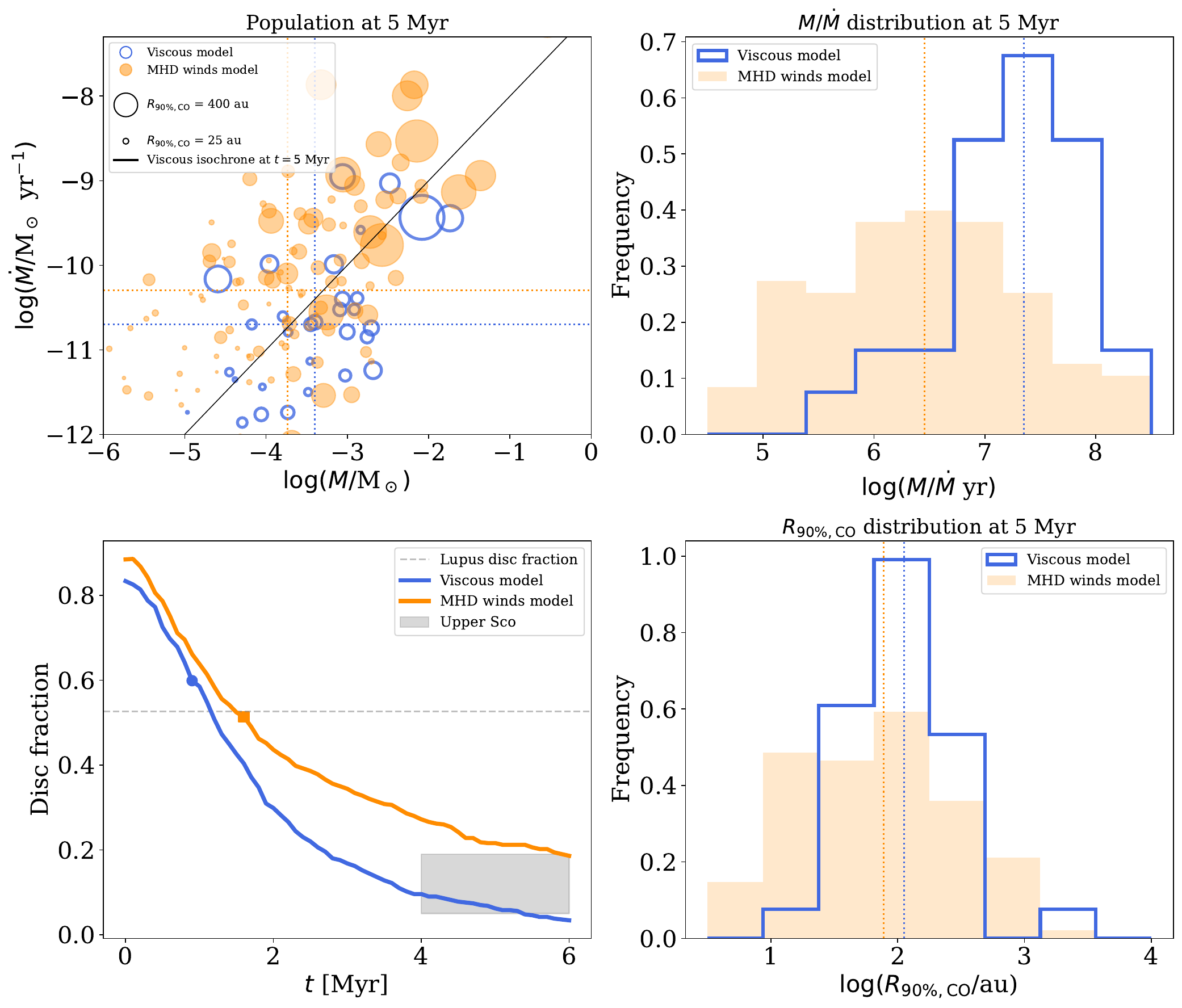} 
     
    \caption{Comparison between the synthetic populations, generated with the best-fit parameters of the viscous-photoevaporative model (blue) and MHD winds model (orange), evolved until $\langle t \rangle$ = 5 Myr. The top-left panel shows the distribution in the $\dot{M}-M$ plane, compared with the viscous isochrone at 5 Myr. The dotted vertical and horizontal lines indicate the median values of the $M$ and $\dot{M}$ distributions, respectively. The size of the circles traces the synthetically computed $R_{90\%, \text{CO}}$ of the discs. The top-right and bottom right panels show the 1D distribution of the $M/\dot{M}$ ratio and $R_{90\%, \text{CO}}$, respectively, where the median values are shown as dotted vertical lines. The bottom-left panel shows the fraction of disc-bearing sources in the two synthetic datasets, compared with the observed disc fraction of Lupus (dashed horizontal line) and the one of Upper-Scorpius (grey rectangle) \citep{Ribas2014}. The blue dot and the orange square indicate the synthetic disc fraction at the age at which the two models best reproduce the Lupus demographics.}
    \label{fig:5_Myr}
\end{figure*}

In the previous section, we showed that the two evolutionary frameworks can reproduce the Lupus demographics with a comparable accuracy.
However, this does not suggest that the two scenarios predict an identical evolution with time.
In fact, it is well established that turbulent-driven accretion and purely wind-driven evolution lead to qualitatively different characteristic behaviours (see e.g., \citealt{Tabone2022model}).
To test the predicted evolution over time of the two frameworks, we generated two synthetic populations of 500 discs, using the model parameters that best reproduce the Lupus population.
For the viscous model, we adopted a mean age of $\langle t \rangle = 0.9$ Myr, whereas for the MHD winds model we used $\langle t \rangle = 1.6$ Myr. 
In both cases, we incorporated the age scatter inferred from the fits.
Subsequently, we evolved the two populations until 5 Myr, which approximately corresponds to the mean age of the Upper Scorpius star-forming region \citep{Pecaut2016}.
We added a scatter to the synthetic points to mimic the observational uncertainties, where we adopted the values described in Section \ref{uncertainties}.
Figure \ref{fig:5_Myr} shows a comparison between the two synthetic populations at 5 Myr.
The viscous model (blue) predicts a larger median disc mass and a lower accretion rate (top-left panel) than the MHD winds model (orange), resulting in most of the viscously evolving discs being placed below the viscous isochrone at 5 Myr. This is further highlighted by the top-right panel, which shows that the viscous $M/\dot{M}$ ratio is larger by more than one order of magnitude compared to the MHD winds one.
Such a behaviour is well explained by Equation \eqref{eqn:viscous_timescale}, which shows that the viscous model yields an increasing $M/\dot{M}$ over time.
In addition, we find that the majority of viscous discs undergo inside-out dispersal due to photoevaporation, resulting in a further decrease of the $M/\dot{M}$ ratio at later stages \citep{Rosotti2017, Somigliana2020}.
This dispersal pathway is different from the one found in the best-fit population of \cite{Tabone2025}, who, fixing the mean age to 2 Myr, inferred that outside-in dispersal was required to better match the low observed $M/\dot{M}$.
The bottom-right panel illustrates the $R_{90\%, \text{CO}}$ distributions, highlighting that, at 5 Myr, the size distributions predicted by the two models do not show significant differences.
However, by comparing the bottom-right panel of Figure \ref{fig:5_Myr} to the theoretical $R$ distribution in Figure \ref{fig:fit} (both a and b), we notice that the median size of discs is expected to increase under both a viscous and a wind-driven evolution.
This trend arises because both mechanisms preferentially disperse small discs before extended ones, leading to a survivorship bias (see \citealt{Somigliana2020, Malanga2025, Tabone2025}; \rev{\citealt{Weder2026}}). 
Finally, the bottom-left panel shows a comparison between the fraction of disc-bearing sources in synthetic populations, highlighting that the viscous model produces a steeper decay with time.
However, this can be attributed to the fact that the viscous model parameters are tuned to reproduce the observed disc fraction at 0.9 Myr, which is lower than the mean age associated with the wind-driven evolution.
As a consequence, this leads to a significantly lower disc fraction at 5 Myr in the viscous model.
\rev{For both models, the disc fraction is not 1 at 0 Myr due to the intrinsic age scatter.}

The comparison between the two evolved best-fit synthetic populations provides a notable result: although the two models reproduce the observed Lupus demographics with similar accuracy, their predictions on disc evolution are substantially different.
Hence, applying \texttt{DEMOS} to multiple star-forming regions will be impactful in our understanding of protoplanetary disc evolution.
Observations of Upper Scorpius \citep{Manara2020, Empey2026} qualitatively show a better compatibility with the expected evolution of MHD winds-driven discs, although an analysis of the complete sample with \texttt{DEMOS} is needed to draw definitive conclusions.
\rev{Moreover, since Upper Scorpius is mildly irradiated \citep{Anania2025}, external photoevaporation should not be neglected.}
The differences between the competing models would become even more pronounced if Upper Scorpius is $\sim$ 10 Myr old, as suggested by \cite{Luhman&Esplin2020}.

\subsection{The legacy of \texttt{DEMOS}}

The application of \texttt{DEMOS}, a Bayesian inference tool described in Paper I, to the Lupus demographics has unveiled precious information about disc evolution models.
Indeed, we optimally exploited the distribution of global disc properties, disc mass $M$, accretion rate $\dot{M}$ and radius $R_{90\%,\text{CO}}$ and the stellar mass $M_\star$, and broadly explored the parameter space.
This allowed us to determine with significant precision the initial conditions and model parameters that most accurately reproduce the observations, under both viscous-photoevaporative and MHD winds-driven evolution.
The accuracy with which the two frameworks match the observed dataset is comparable, which does not allow us to disentangle disc evolution with solely the Lupus demographics.
Nevertheless, despite both models reproducing the observations at the age of Lupus, they predict a substantially different evolution at later ages.
One of the main discrepancies lies in the initial size distribution (shown in Figure \ref{fig:initial_conditions}).
Hence, comparing the initial conditions yielded by \texttt{DEMOS} for both models to the expected \citep{Bate2018, Lebreuilly2024} and observed (\citealt{Tychoniec2018, Maury2019, Tobin2020}, see also \citealt{Maury2022} for a review) Class 0/I properties will be fundamental to assess which accretion mechanism is more likely to be dominant in protoplanetary discs.
In addition, the properties of the two best-fit populations are significantly different when evolved until 5 Myr (see Section \ref{evolution}).
Consequently, the application of \texttt{DEMOS} to an older population, such as Upper Scorpius, will break the degeneracy, allowing us to rule out one of the two models.
\rev{Future works will need to allow \texttt{DEMOS} to tackle selection and observational biases, fundamental properties of many existing disc samples with a lower degree of completeness than Lupus.}

\subsection{The importance of upper limits in disc size}
\label{R_upper_limits}
\begin{figure}[h!]
    \centering
    \includegraphics[width=0.9\linewidth]{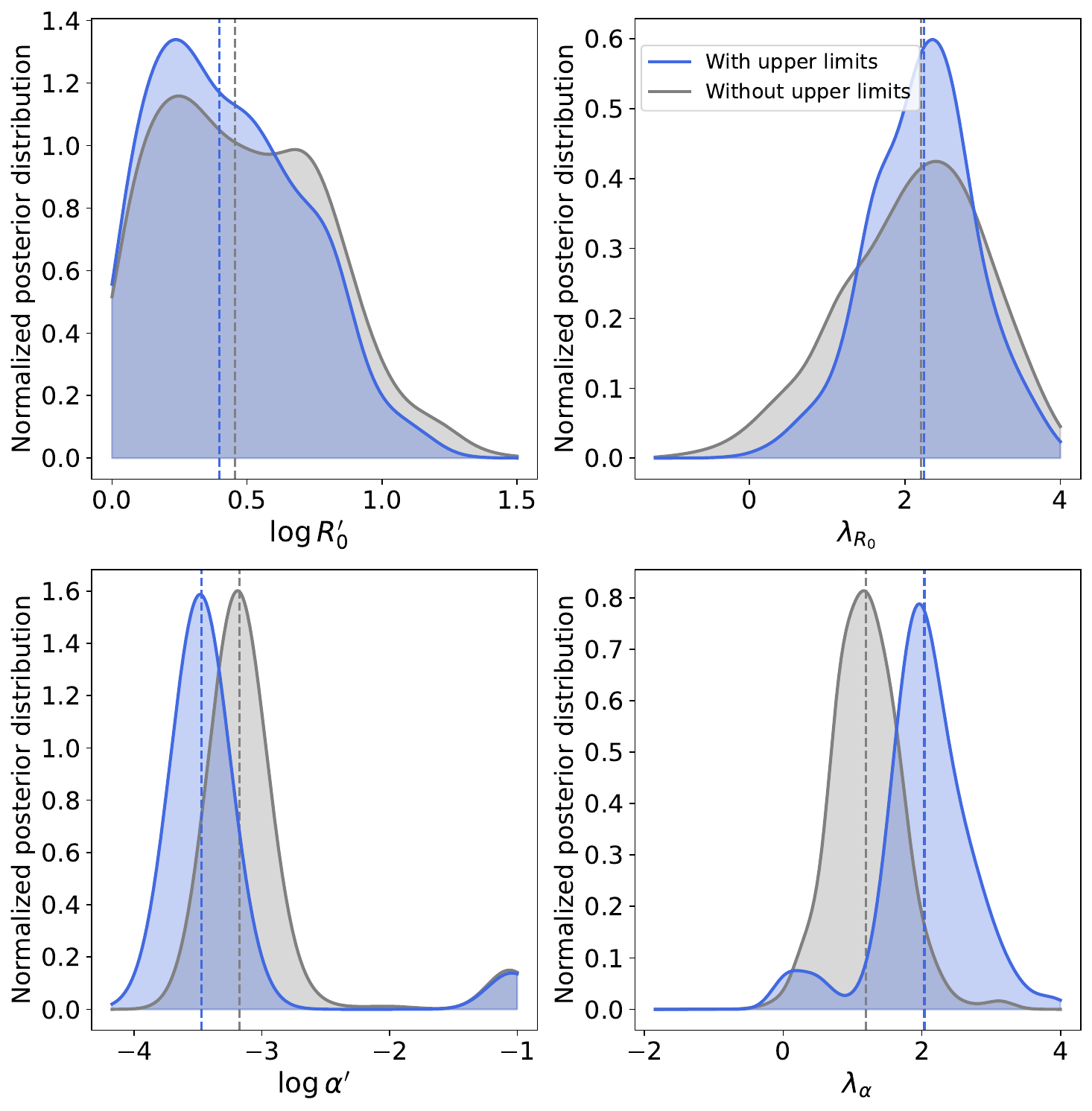} 
     
    \caption{Comparison between 1D posterior distributions of $\log R_0'$, $\lambda_{R_0}$, $\log \alpha '$ and $\lambda_\alpha'$, for the viscous-photoevaporative model, inferred with (blue) and without (grey) the inclusion of upper limits in $R_{90\%,\text {CO}}$ arising from $^{12}$CO non-detections. The medians of the distributions are indicated by vertical dashed lines.}
    \label{fig:R_upper_limits}
\end{figure}

$^{12}$CO non-detections play a crucial role in our inference.
We considered an upper limit on $R_{90\%, \text{CO}}$ for sources where $^{12}$CO is not detected, following \cite{Miotello2021} and \cite{Ricciardi2026}, who explain low CO fluxes and non-detections with compact discs models. 
This choice was motivated by the goal of fully exploiting the information encapsulated in the demographics, allowing a higher accuracy and precision in the inference.
However, this assumption had a significant impact on our estimates of the initial disc radius distribution $R_0$ and on the angular momentum efficiency parameter $\alpha$.
The influence of size upper limits is also highlighted by the comparison between the bets-fit and observed size distribution (see Figure \ref{fig:fit}).
Figure \ref{fig:R_upper_limits} shows the 1D posterior distribution of four parameters of the viscous model, $\log R_0'$, $\lambda_{R_0}$, $\log \alpha '$, and $\lambda_\alpha'$, inferred using the complete dataset (blue) and without considering the upper limits in $R_{90\%,\text{CO}}$ due to non-detections (grey).
\rev{In the latter case, sources whose disc size is not measured are still included in the analysis, but are assigned a non-informative upper limit of 10000 au on $R_{90\%, \text{CO}}$.}
The Figure highlights that excluding upper limits slightly increases the inferred $\log R_0'$ and increases $\log \alpha'$ by a factor 2. 
This occurs because, when upper limits are not included, \texttt{DEMOS} calibrates synthetic populations solely on the available information on disc sizes, which is represented by the subset of resolved, more extended discs.
Moreover, the $\alpha-M_\star$ correlation flattens when upper limits are excluded, because most of the discs where $^{12}$CO is undetected surround a low mass star (see Figure \ref{fig:Lupus_demographics}).
Consequently, the inclusion of upper limits primarily affects the $\alpha$ of discs around low-mass stars, while the models predict $\alpha \sim 3\times10^{-3}$ for discs around 1 M$_\odot$ stars regardless of whether upper limits are included.
We also compared the posterior distributions obtained using the complete dataset with those derived after excluding upper limits and the subset of small discs whose gas sizes were measured by \cite{Ricciardi2026}. 
In this case, we find a higher value of $\log R_0'$ and a less steep $R_0-M_\star$ correlation.
This comparison sheds light on the critical importance of size measurements in disc populations and motivates further observational efforts to expand the sample of sizes across multiple star-forming regions.

\subsection{Low estimated ages: a hint of a different evolution scenario?}

For both evolutionary models, we estimated the mean age at which synthetic populations are best reproducing the data. 
Since constraints on the age of the Lupus population exist in the literature \citep{Alcala2017, Galli2020_Lupus}, we adopted a prior distribution for $\langle t \rangle$, centred at 2 Myr.
This choice reflects the assumption that the age of a Young Stellar Object (YSO) can be approximated by the length of the disc-phase, because of the relative briefness protostellar stage. 
However, the discrepancy between typical age estimates and our results, especially for the viscous model, could challenge the validity of such assumption.
In fact, the age inferred by \texttt{DEMOS} could be interpreted as the duration of the phase during which YSOs can be modeled as isolated discs.
Prior to such epoch, a phase of non-negligible duration could occur, in which environmental effects shape the evolution of the YSO, and set the initial conditions for the subsequent disc phase.
We foresee that an application of \texttt{DEMOS} to different star-forming regions will assess the impact of environments on the initial conditions.

Moreover, disc evolution models that include environmental effects should be explored.
In particular, the evidence of spatially correlated accretion rate in Lupus suggests that late-infall may play an important role in disc evolution \citep{Winter2024}.
Since late-infall is expected to decrease the $M/\dot{M}$ ratio \citep{Winter2024a}, we expect that this mechanism could shift the inferred mean disc ages toward values more consistent with those derived from the HR diagram.
This possibility is especially intriguing for Chamaeleon I, which has $\langle M/\dot{M} \rangle \sim$ 0.5 Myr \citep{Pascucci2016, Manara_2017, Mulders2017}, about 4 times lower than in Lupus, despite having a similar estimated age \citep{Galli2021}.
Future works should also explore the impact of different assumptions regarding the gas-to-dust ratio, which is expected to evolve with age (e.g., \citealt{Sellek2020}).

\section{Conclusions}

In this paper, we applied the Bayesian inference tool \texttt{DEMOS}, presented in Malanga et al. (2026a), to the Lupus disc population. 
We considered as observables the stellar mass $M_\star$, the disc mass $M$, the accretion rate $\dot{M}$, the disc size $R_{90\%, \text{CO}}$, and the fraction of disc-bearing sources in the population.
In our inference, we did not consider confirmed multiple systems, resulting in a sample of 89 sources.
We obtained the posterior distribution of model parameters and initial conditions that most accurately reproduce the Lupus demographics for both the viscous-photoevaporative framework and the MHD winds model.
Our results can be summarised as follows:
\begin{enumerate}
    \item The Lupus demographics are best reproduced by an initial disc population where the bulk properties $M_0$ and $R_0$ scale positively with the stellar mass. Such correlations should somehow be injected by a mechanism prior to the disc-phase.
    \item Both models predict a strong scaling of initial disc size with stellar masses, with very compact discs (with sizes $\sim  3-5$ au) around $0.3\ \text{M}_\odot$ stars and more extended discs $\sim  20-40$ au) around 1 M$_\odot$ stars. This may have significant impact on the planet formation potential of M dwarfs compared to solar mass stars.
    For the turbulence-driven scenario, we expect a steeper correlation between and $R_0$ and $M_\star$ and a larger intrinsic spread.
    These results need to be compared to observations and theoretical predictions of Class 0/I objects.
    \item We find that the typical angular momentum transport parameter $\alpha$ is $\sim 3-4 \times 10^{-4}$ for both models, in agreement with recent estimates in the literature. Moreover, we find that it scales with the stellar mass following an almost quadratic relation.
    \item We derive that the typical wind's mass-loss rate \rev{due to internal photoevaporation} is $\sim  10^{-9}$ M$_\odot$ yr$^{-1}$, consistent with simulations, and that it is steeply correlated with $M_\star$ with an exponent $\sim 2.5$.
    \item Both models predict lower average ages for the Lupus population compared to the expected stellar age (0.9 and 1.6 Myr for the viscous and wind-driven case, respectively), with a significant discrepancy especially for the viscous model. This is due to the necessity to reproduce observed discs with a low $M/\dot{M}$ ratio. 
    \item Despite a slight preference for the viscous-photoevaporative model, the inference of the Lupus demographics does not allow us to rule out either of the two evolutionary models.
    However, the viscous model is less compatible with the data if the mean age is fixed to 2 Myr, in agreement with \cite{Tabone2025}.
    \item The two best-fit synthetic populations exhibit systematically different behaviours when evolved until 5 Myr, suggesting that applying \texttt{DEMOS} to a more evolved region, such as Upper Scorpius, will provide key insights into the mechanisms driving disc evolution.
\end{enumerate}
In addition, in this work we presented the first application of \texttt{DEMOS} to an observed population.
We showed that by exploiting the whole distribution in the observable space, we can put significant constraints on complex models with many free parameters, such as the ones which expect correlations between initial conditions and the stellar mass.
Moreover, the statistical robustness of the framework allows to obtain the probability distribution of model parameters.
The application of this tool to various star-forming regions will improve our understandings on the evolution of protoplanetary discs.

\begin{acknowledgements}

We thank the anonymous referee for their useful comments that helped us improve the clarity of this manuscript.
We thank Marco Lombardi for the insightful discussion about Bayesian statistics, which improved the clarity and correctness of the statistical analysis.
The authors acknowledge support from the European Union (ERC Starting Grant DiscEvol, project number 101039651 and ERC, WANDA, project number 101039452), from Fondazione Cariplo, grant No. 2022-1217, from the European Union’s Horizon 2020 research and innovation programme under the Marie Sklodowska-Curie grant agreement No 823823 (Dustbusters RISE project), and from the ERC Synergy Grant “ECOGAL” (project ID 855130). 
GL acknowledges financial contribution from PRIN-MUR 20228JPA3A.
We acknowledge financial support from the project "Space Observations of Proto-planetary Accretion Disc Evolution (SPADE)” in the framework of the "PROgramma di RIcerca Spaziale di base (PRORIS)", financed by the Italian Ministry of University and Research (Ministero dell’Università e della Ricerca; MUR).
Views and opinions expressed are, however, those of the authors only and do not necessarily reflect those of the European Union or the European Research Council. Neither the European Union nor the granting authority can be held responsible for them.

\end{acknowledgements}

\bibliographystyle{aa}

\bibliography{repbib}

\clearpage
\onecolumn
\begin{appendix}

\section{The complete sample}
Table \ref{tab:Lupus} shows the entirety of the Lupus demographics considered in our analysis. The first column lists the index of each source, used to identify the objects in Figure \ref{fig:Lupus_demographics}. 
The second column lists the source name, while the following columns show the logarithm of the four considered observables: stellar mass, disc mass, accretion rate and disc size.
Values preceded by a "<" symbol denote upper limits. 
\begin{center}
\begin{longtable}{r l c c c c}  
\caption{Sample within the Lupus demographics considered in this analysis.} \\
\hline
Index & Source name & $\log M_\star$ & $\log M$ & $\log \dot{M}$ & $ \log R_{90\%\mathrm{CO}} $ \\
\hline
\endfirsthead
\multicolumn{6}{c}{{\tablename\ \thetable{} -- continued}} \\
\hline
Index & Source name & $\log M_\star$ & $\log M$ & $\log \dot{M}$ & $ \log R_{90\%\mathrm{CO}} $ \\
\hline
\endhead
\hline \multicolumn{6}{r}{{Continued on next page}} \\
\endfoot
\hline
\endlastfoot
1 & Sz65 & $-0.21$ & $-2.20$ & $<$$-9.48$ & $2.22$$^{(3)}$ \\
2 & Sz66 & $-0.54$ & $-2.85$ & $-8.51$ & $1.88$$^{(3)}$ \\
3 & J15430131-3409153 & $<$$0.50$ & $<$$-4.22$ & $<$$-7.00$ & $<$$1.86$$^{(4)}$ \\
4 & J15430227-3444059 & $<$$0.50$ & $<$$-4.22$ & $<$$-7.00$ & $<$$1.65$$^{(4)}$ \\
5 & J15445789-3423392 & $-1.06$ & $<$$-4.25$ & $-10.68$ & $<$$1.65$$^{(4)}$ \\
6 & J15450634-3417378 & $<$$0.50$ & $-2.86$ & $<$$-7.00$ & $<$$4.00$ \\
7 & J15450887-3417333 & $-1.04$ & $-2.35$ & $-8.16$ & $<$$4.00$ \\
8 & Sz69 & $-0.82$ & $-2.77$ & $-9.34$ & $2.09$$^{(1)}$ \\
9 & Sz71 & $-0.38$ & $-1.83$ & $-9.03$ & $2.50$$^{(3)}$ \\
10 & Sz72 & $-0.43$ & $-2.91$ & $-8.60$ & $1.47$$^{(3)}$ \\
11 & Sz73 & $-0.15$ & $-2.62$ & $-8.05$ & $2.03$$^{(1)}$ \\
12 & Sz76 & $-0.73$ & $-2.99$ & $-9.11$ & $2.15$$^{(1)}$ \\
13 & Sz81A & $-0.73$ & $-3.02$ & $-8.89$ & $<$$1.66$$^{(4)}$ \\
14 & J15560210-3655282 & $<$$0.50$ & $-2.28$ & $<$$-7.00$ & $2.08$$^{(1)}$ \\
15 & Sz82 / IM Lup & $-0.14$ & $-1.39$ & $-7.85$ & $2.93$$^{(1)}$ \\
16 & Sz83 / RU Lup & $-0.26$ & $-1.43$ & $-7.00$ & $2.51$$^{(1)}$ \\
17 & Sz84 & $-0.83$ & $-2.55$ & $-9.08$ & $2.29$$^{(1)}$ \\
18 & Sz129 & $-0.14$ & $-1.76$ & $-8.27$ & $2.24$$^{(1)}$ \\
19 & J15592523-4235066 & $-1.07$ & $-3.98$ & $-11.02$ & $<$$1.65$$^{(4)}$ \\
20 & RYLup & $0.10$ & $-1.71$ & $-8.05$ & $2.41$$^{(1)}$ \\
21 & J16000060-4221567 & $-0.83$ & $-3.66$ & $-9.58$ & $<$$1.32$$^{(2)}$ \\
22 & J16000236-4222145 & $-0.72$ & $-1.94$ & $-9.48$ & $2.42$$^{(1)}$ \\
23 & J16002612-4153553 & $-0.97$ & $-3.99$ & $-9.62$ & $1.68$$^{(2)}$ \\
24 & Sz130 & $-0.40$ & $-3.35$ & $-9.10$ & $2.08$$^{(1)}$ \\
25 & MYLup & $0.08$ & $-1.82$ & $-8.00$ & $2.26$$^{(1)}$ \\
26 & Sz131 & $-0.52$ & $-3.11$ & $-9.17$ & $1.86$$^{(3)}$ \\
27 & J16011549-4152351 & $<$$0.50$ & $-2.31$ & $<$$-7.00$ & $<$$4.00$ \\
28 & EXLup & $-0.24$ & $-2.38$ & $-7.81$ & $<$$2.20$$^{(1)}$ \\
29 & Sz133 & $<$$0.50$ & $-2.22$ & $<$$-7.00$ & $2.31$$^{(1)}$ \\
30 & J16070384-3911113 & $<$$0.50$ & $-3.66$ & $<$$-7.00$ & $2.37$$^{(1)}$ \\
31 & J16070854-3914075 & $<$$0.50$ & $-2.04$ & $<$$-7.00$ & $2.53$$^{(1)}$ \\
32 & Sz90 & $-0.14$ & $-2.69$ & $-8.93$ & $1.88$$^{(1)}$ \\
33 & Sz91 & $-0.28$ & $-2.66$ & $-9.08$ & $2.49$$^{(1)}$ \\
34 & J16073773-3921388 & $-1.17$ & $-3.88$ & $-10.19$ & $<$$1.69$$^{(4)}$ \\
35 & Sz95 & $-0.54$ & $-3.38$ & $-9.38$ & $<$$1.64$$^{(4)}$ \\
36 & J16075475-3915446 & $<$$0.50$ & $-3.96$ & $<$$-7.00$ & $<$$1.63$$^{(4)}$ \\
37 & J16080017-3902595 & $-1.11$ & $-3.54$ & $-10.27$ & $<$$1.65$$^{(4)}$ \\
38 & J16080175-3912316 & $<$$0.50$ & $<$$-4.22$ & $<$$-7.00$ & $<$$1.64$$^{(4)}$ \\
39 & Sz96 & $-0.33$ & $-3.51$ & $-9.39$ & $1.52$$^{(1)}$ \\
40 & J16081497-3857145 & $-1.19$ & $-3.16$ & $-10.28$ & $1.80$$^{(2)}$ \\
41 & Sz97 & $-0.71$ & $-3.36$ & $-9.77$ & $1.26$$^{(2)}$ \\
42 & Sz98 & $-0.26$ & $-1.64$ & $-7.44$ & $<$$4.00$ \\
43 & Sz99 & $-0.74$ & $<$$-4.22$ & $-9.61$ & $<$$1.65$$^{(4)}$ \\
44 & Sz100 & $-0.89$ & $-2.31$ & $-9.61$ & $2.11$$^{(1)}$ \\
45 & J160828.1-391310 & $-1.15$ & $<$$-4.16$ & $<$$-11.32$ & $<$$1.65$$^{(4)}$ \\
46 & Sz102 & $<$$0.50$ & $-2.86$ & $<$$-7.00$ & $1.87$$^{(1)}$ \\
47 & Sz103 & $-0.71$ & $-2.98$ & $-9.25$ & $1.35$$^{(2)}$ \\
48 & J16083070-3828268 & $0.10$ & $-2.06$ & $<$$-9.20$ & $2.49$$^{(1)}$ \\
49 & Sz104 & $-0.97$ & $-3.61$ & $-9.84$ & $<$$1.30$$^{(2)}$ \\
50 & J160831.1-385600 & $<$$0.50$ & $<$$-4.22$ & $<$$-7.00$ & $<$$1.65$$^{(4)}$ \\
51 & V1094Sco & $-0.19$ & $-1.39$ & $-7.88$ & $2.92$$^{(3)}$ \\
52 & Sz106 & $-0.26$ & $-3.79$ & $-10.09$ & $<$$1.45$$^{(2)}$ \\
53 & Sz108B & $-0.91$ & $-2.56$ & $-9.51$ & $<$$1.68$$^{(4)}$ \\
54 & V1192Sco & $<$$0.50$ & $-4.16$ & $<$$-7.00$ & $<$$2.32$$^{(1)}$ \\
55 & Sz110 & $-0.72$ & $-2.79$ & $-8.75$ & $1.25$$^{(2)}$ \\
56 & J16085324-3914401 & $-0.54$ & $-2.72$ & $-10.03$ & $1.74$$^{(3)}$ \\
57 & J16085373-3914367 & $-1.17$ & $-3.69$ & $-10.81$ & $<$$1.65$$^{(4)}$ \\
58 & Sz111 & $-0.28$ & $-1.87$ & $-9.47$ & $2.67$$^{(1)}$ \\
59 & J16085529-3848481 & $-1.15$ & $-3.81$ & $-10.60$ & $<$$1.44$$^{(2)}$ \\
60 & Sz112 & $-0.87$ & $-3.48$ & $-9.82$ & $1.60$$^{(2)}$ \\
61 & Sz113 & $-0.87$ & $-2.62$ & $-8.99$ & $1.44$$^{(2)}$ \\
62 & J16085828-3907355 & $<$$0.50$ & $<$$-4.22$ & $<$$-7.00$ & $<$$1.65$$^{(4)}$ \\
63 & J16085834-3907491 & $<$$0.50$ & $<$$-4.22$ & $<$$-7.00$ & $<$$1.65$$^{(4)}$ \\
64 & J16090141-3925119 & $-0.70$ & $-2.90$ & $-9.89$ & $2.29$$^{(1)}$ \\
65 & Sz114 & $-0.80$ & $-2.04$ & $-9.10$ & $2.23$$^{(1)}$ \\
66 & Sz115 & $-0.82$ & $<$$-4.22$ & $-9.45$ & $<$$1.64$$^{(4)}$ \\
67 & J16091644-3904438 & $<$$0.50$ & $<$$-4.16$ & $<$$-7.00$ & $<$$1.66$$^{(4)}$ \\
68 & J16092032-3904015 & $<$$0.50$ & $<$$-4.22$ & $<$$-7.00$ & $<$$1.38$$^{(4)}$ \\
69 & J16092317-3904074 & $<$$0.50$ & $<$$-4.22$ & $<$$-7.00$ & $<$$1.65$$^{(4)}$ \\
70 & J16092697-3836269 & $-0.83$ & $-3.40$ & $-8.11$ & $2.05$$^{(1)}$ \\
71 & J160934.2-391513 & $<$$0.50$ & $<$$-4.22$ & $<$$-7.00$ & $<$$1.65$$^{(4)}$ \\
72 & J16093928-3904316 & $<$$0.50$ & $<$$-4.22$ & $<$$-7.00$ & $<$$1.65$$^{(4)}$ \\
73 & Sz117 & $-0.63$ & $-3.03$ & $-8.88$ & $1.43$$^{(2)}$ \\
74 & Sz118 & $-0.08$ & $-2.26$ & $-9.11$ & $2.17$$^{(1)}$ \\
75 & J16100133-3906449 & $-1.00$ & $<$$-4.08$ & $-9.62$ & $<$$1.73$$^{(4)}$ \\
76 & J16101307-3846165 & $<$$0.50$ & $<$$-4.29$ & $<$$-7.00$ & $<$$1.64$$^{(4)}$ \\
77 & J16101857-3836125 & $-1.02$ & $<$$-4.20$ & $-10.49$ & $<$$1.65$$^{(4)}$ \\
78 & J16101984-3836065 & $-1.21$ & $<$$-4.22$ & $-10.35$ & $<$$1.64$$^{(4)}$ \\
79 & J16102741-3902299 & $<$$0.50$ & $<$$-4.22$ & $<$$-7.00$ & $<$$1.65$$^{(4)}$ \\
80 & J16102955-3922144 & $-0.82$ & $-3.17$ & $-9.92$ & $2.13$$^{(1)}$ \\
81 & J16115979-3823383 & $<$$0.50$ & $<$$-4.20$ & $<$$-7.00$ & $<$$1.65$$^{(4)}$ \\
82 & J16120445-3809589 & $<$$0.50$ & $<$$-4.22$ & $<$$-7.00$ & $<$$1.65$$^{(4)}$ \\
83 & J16124373-3815031 & $-0.33$ & $-2.57$ & $-9.08$ & $2.04$$^{(3)}$ \\
84 & J16134410-3736462 & $-1.02$ & $-3.81$ & $-9.01$ & $<$$1.37$$^{(2)}$ \\
85 & AKC2006-18 & $-1.36$ & $<$$-4.75$ & $-10.99$ & $<$$1.65$$^{(4)}$ \\
86 & J154518.5-342125 & $-1.21$ & $-3.39$ & $-10.54$ & $1.56$$^{(1)}$ \\
87 & Lup706 & $-1.37$ & $-4.14$ & $-11.63$ & $<$$1.94$$^{(1)}$ \\
88 & SONYC-Lup3-7 & $-1.64$ & $-4.41$ & $-10.88$ & $<$$1.56$$^{(1)}$ \\
89 & SONYC-Lup3-10 & $-1.64$ & $<$$-4.75$ & $-10.53$ & $<$$1.65$$^{(4)}$ \\
\caption{\textbf{Note:} The symbol "<" preceding values indicates an upper limit.
$^{(1)}$ $R_{90\%, \text{CO}}$ measures are from \cite{Sanchis2021}.
$^{(2)}$ $R_{90\%, \text{CO}}$ measures are from \cite{Ricciardi2026}.
$^{(3)}$ $R_{90\%, \text{CO}}$ measures are from \cite{Trapman2025_AGE_PRO_sizes}.
$^{(4)}$ $R_{90\%, \text{CO}}$ is obtained with the flux-radius relation \citep{Zagaria2023b}.}
\label{tab:Lupus}
\end{longtable}
\end{center}
\section{Results of \texttt{emcee} and \texttt{UltraNest} analyses}
\label{Appendix_results}
\begin{figure*}[ht!]
    \centering
    \includegraphics[width=\linewidth]{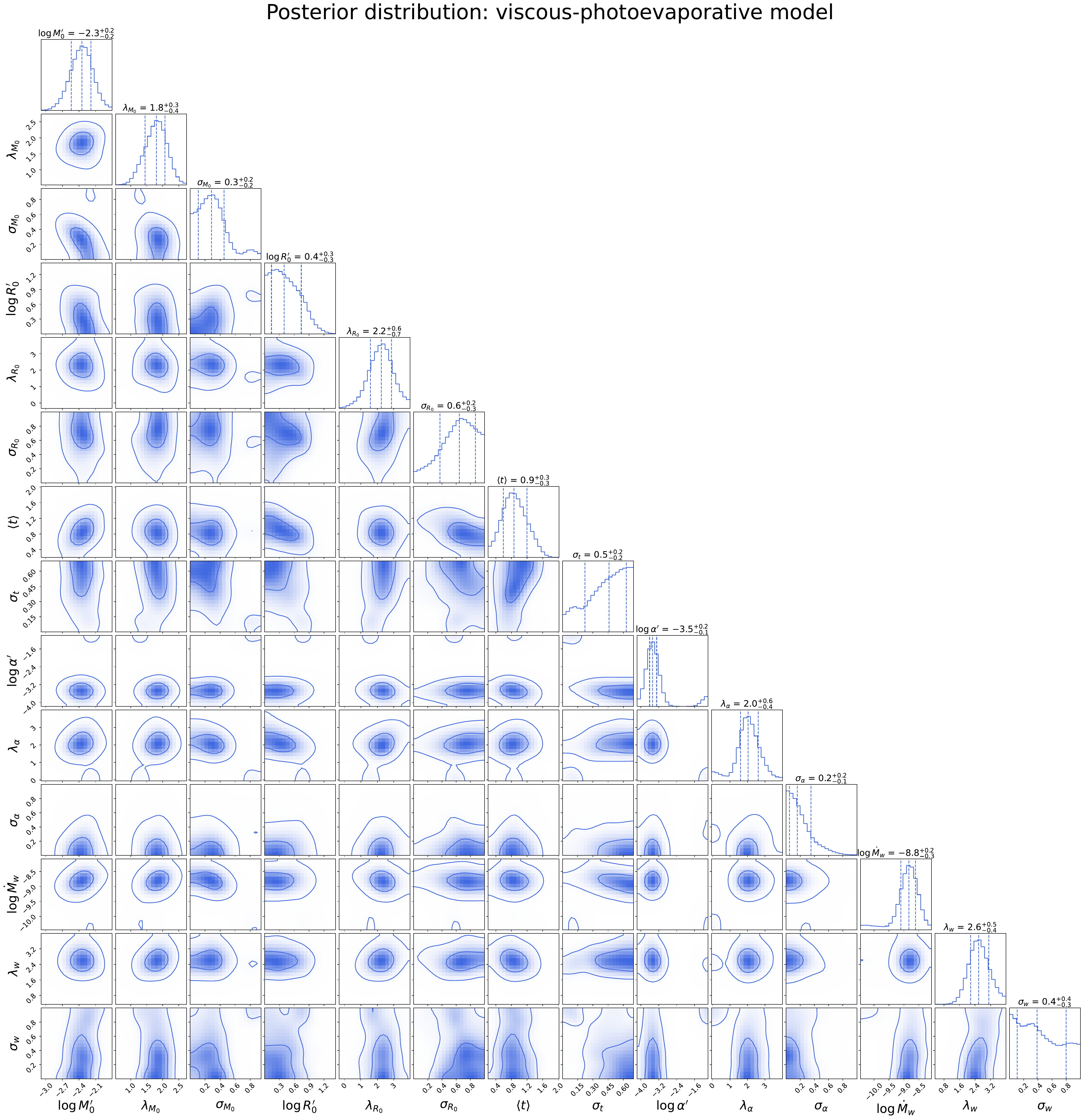} 
     
    \caption{\small{Corner plot showing the posterior distribution of the free parameters of the viscous-photoevaporative model. The contours in 2D distributions show the 1$\sigma$ and 2$\sigma$ confidence intervals. The three vertical dashed lines in the histograms mark the 16$^\text{th}$, 50$^\text{th}$ and 84$^\text{th}$ percentiles.}}
    \label{fig:Cornerplot_viscous}
\end{figure*}

\begin{figure*}[ht!]
    \centering
    \includegraphics[width=\linewidth]{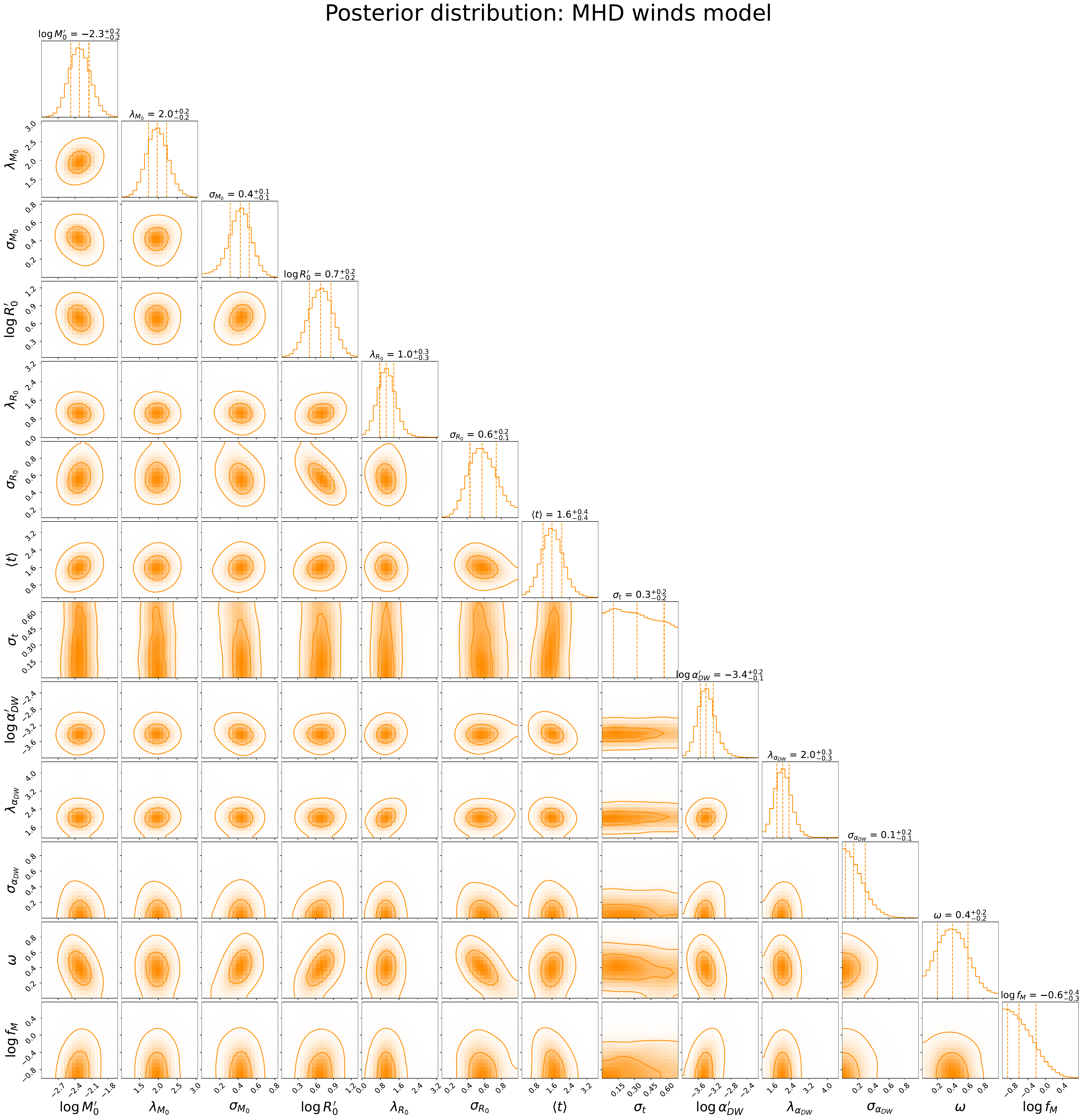} 
     
    \caption{\small{Corner plot showing the posterior distribution of the free parameters of the MHD winds model. The contours in 2D distributions show the 1$\sigma$ and 2$\sigma$ confidence intervals. The three vertical dashed lines in the histograms mark the 16$^\text{th}$, 50$^\text{th}$ and 84$^\text{th}$ percentiles.}}
    \label{fig:Cornerplot_MHD}
\end{figure*}

Figures \ref{fig:Cornerplot_viscous} and \ref{fig:Cornerplot_MHD} display the posterior distributions of the free parameters of the viscous-photoevaporative framework and wind-driven model, respectively, obtained with \texttt{emcee}.
The plots show that no parameters are particularly correlated.
Moreover, while in the MHD winds scenario all parameters except for the age scatter, are constrained, the viscous model poorly constrains the spread of the $\dot{M}_\text{w}$ distribution and the spread of the $R_0$ distribution.
The main challenge that we encountered performing these fits has been the stochastic nature of the likelihood.
In fact, at each evaluation, a new synthetic population is generated, following the distribution generated by the free parameters $\boldsymbol{\theta}$.
This causes the likelihood to be slightly stochastic and not-well defined for a set of free parameters.
Such effects results in a low \texttt{emcee} acceptance fraction, typically between 1\% and 5\%, and in slightly broader posterior distributions.
An alternative approach would be to fix the random seed in the Monte Carlo population generation. However, to suppress statistical fluctuations under this setup, the size of the synthetic populations would need to be increased to at least $\sim$1000 discs, doubling the computational cost of the simulations.
However, we performed multiple independent fits for both models and consistently recovered the same posterior distribution.
This validates the statistical robustness of our parameter estimation.

In addition, we performed the inference using the nested sampling algorithm \texttt{UltraNest} \citep{Buchner2021}.
This method, unlike MCMC sampling, which explores the parameters space targeting regions of high likelihood, evaluates the integral of the likelihood over the prior, calculating the evidence.
The process terminates as soon as the estimated evidence reaches a predefined level of precision.
However, the efficiency of nested-sampling notoriously decreases with a number of free parameters $k>10$ \citep{Buchner2021}.
This caused none of our simulations to converge, although the threshold that we selected, of $\Delta \ln Z = 2.5$, was relatively loose.
Due to the heavy computational cost, we had to interrupt the inference at $\sim 95\%$ of integral calculation for the MHD winds model and $\sim 30\%$ for the viscous model, which is 5 to 8 times more expensive due to internal photoevaporation.
This does not have a strong impact on posterior distributions, which exquisitely overlap with those obtained with \texttt{emcee}.
In contrast, we could not derive robust uncertainty estimates on the evidence for both models.
While we expect $\Delta \ln Z$ not to be significantly larger than 2.5 for the MHD winds model, no robust constraint can be established for the viscous model.
Nevertheless, the similarity between the two estimated evidences suggests that the evidence difference is not sufficient to claim a strong preference for either model.
In order to allow a more precise evidence computation, the population synthesis generation process must be further optimized (\rev{e.g., by using neural networks to learn the mapping from model parameters to the likelihood}).

\section{Disc fraction per $M_\star$ bin}
\label{disc_fraction_stellar_mass}

\begin{figure}[ht!]
    \centering
    \includegraphics[width=0.7\linewidth]{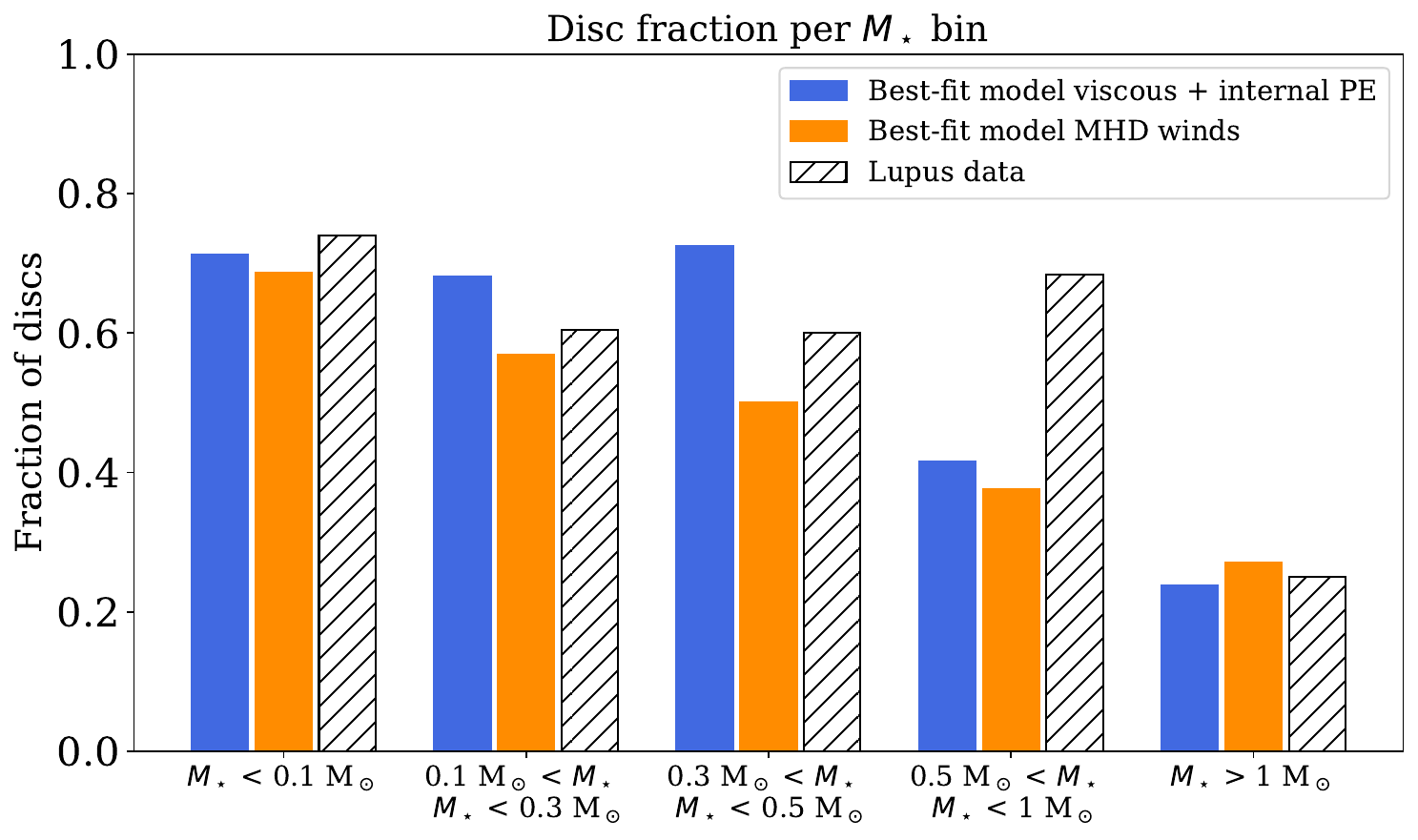} 
     
    \caption{\small{Bar diagram comparing the disc fraction is six $M_\star$ bins of Lupus (white hatched), and synthetic population, for the viscous (blue) and MHD winds model (orange), respectively.}}
    \label{fig:disc_fraction_mstar}
\end{figure}

The dependence of the disc fraction on the stellar mass has a strong impact on planet formation, because it identifies the systems where giant planets have more time to form.
Such quantity is recovered by \texttt{DEMOS} by fitting for the $M_\star$ distribution of the disc-bearing sources, because it shows the $M_\star$ probability density with respect to the initial mass function.
Figure \ref{fig:disc_fraction_mstar} shows a comparison between the observed (white hatched) and synthetic disc fraction, for both the viscous (blue) and MHD winds (orange) models.
The fraction is computed in six $M_\star$ bins, by dividing the number of disc-bearing objects in each bin by the total number of objects within the same bin.
The Figure shows a good agreement between the models and observations, confirming that \texttt{DEMOS} is sensitive to the dependence of the disc fraction on $M_\star$.
The only exception is represented by the bin of $M_\star$ within 0.5 M$_\odot$ and 1 M$_\odot$, where the models underestimate the fraction of survived discs.
However, \rev{the statistics for high mass bins are not robust, as only 22 observed sources are distributed across the three highest-mass bins.
In general}, we reproduce the observed trend that disc surrounding low-mass stars are generally longer-lived \citep{Carpenter2006, Yasui2014, Ribas2015, Richert2018, Pfalzner2022}. 
This feature is more pronounced for $M_\star>1$ M$_\odot$, although a weaker trend is noticeable even at lower masses.
The average disc fraction is about $\sim 50-60 \%$, consistent with that of Lupus.

In order to reproduce this trend, \texttt{DEMOS} adjusted the slopes of the $\alpha-M_\star$ and $\dot{M}_\text{w}-M_\star$ correlations for the visocus-photoevaporative model and $\alpha_\text{DW}-M_\star$ for the wind-driven model, which regulate the dispersal timescale as a function of $M_\star$.
Our results for the viscous model deviate from the theoretical expectation that discs around massive stars are longer-lived \citep{Malanga2025}.
This is due to the fact that an approximately linear dependence is usually assumed between $\dot{M}_\text{w}$ and $M_\star$, due to the outcome of numerical simulations \citep{Picogna2021}.
In contrast, we find that the slope of the power law is steeper ($\sim 2.5$).
We stress that this result should not be interpreted as evidence of a physical dependence, but rather as the value that best reproduces the observed Lupus demographics within our framework.
Future works should investigate the dependence of the photoevaporative mass-loss rate on stellar mass, obtained with state-of-the-art hydrodynamical and thermo-chemical simulations, in order to assess the robustness of this finding.

\section{Evolution of the slope of the $\dot{M}- M_\star$ correlation}
\label{slopes}
\begin{figure}[ht!]
    \centering
    \includegraphics[width=0.7\linewidth]{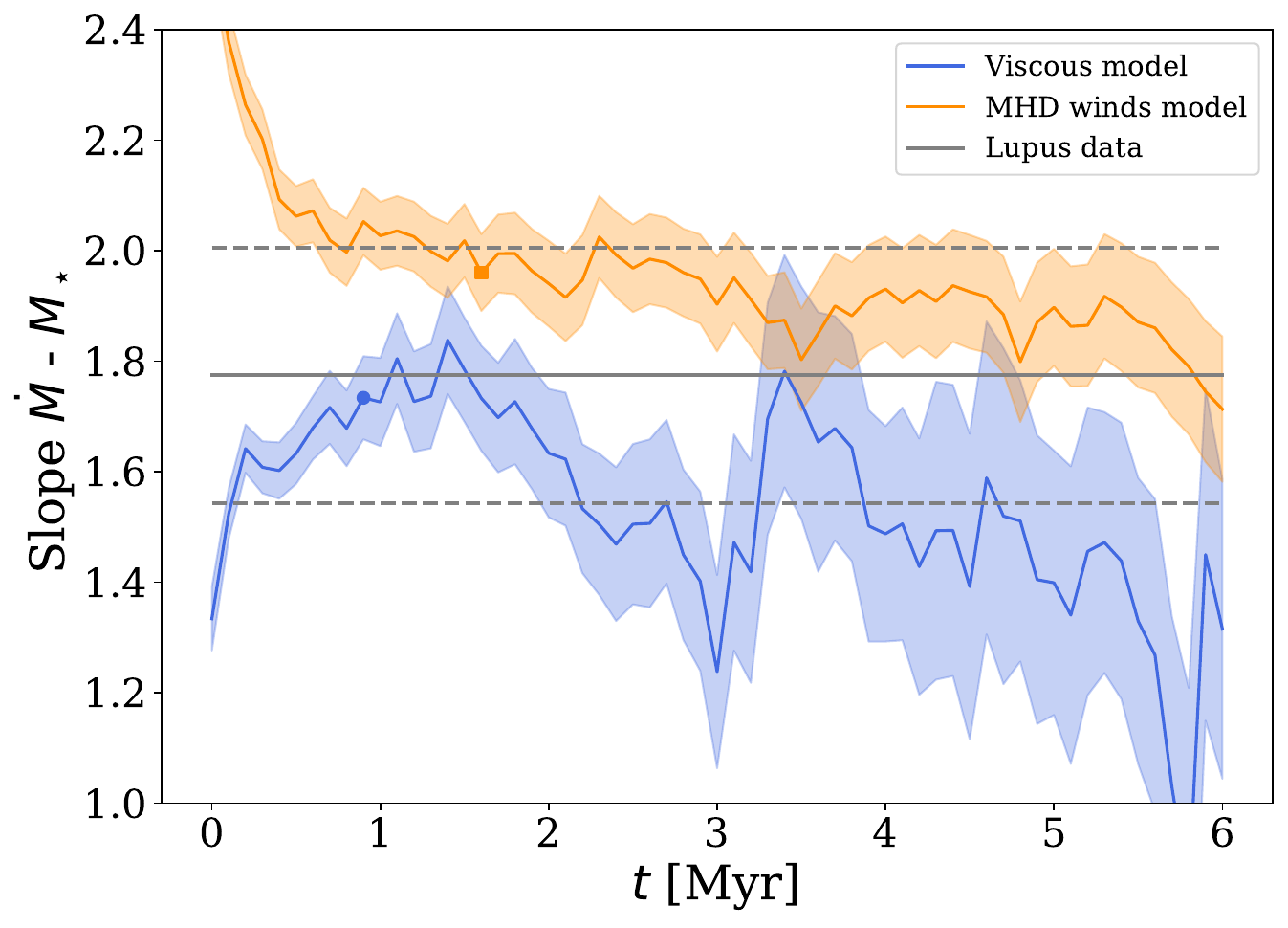} 
     
    \caption{\small{Comparison between the evolution of the $\dot{M}-M_\star$ slope over time predicted by the viscous model (blue) and by the MHD winds model (orange). Solid lines mark the median slope values, while shaded regions represent the $1\sigma$ confidence intervals. The grey horizontal solid line indicates the slope measured from the Lupus data, with dashed lines marking its $1\sigma$ confidence interval. The blue dot and the orange square identify the slope at the timescale at which the models best reproduce Lupus.}}
    \label{fig:slopes}
\end{figure}
Figure \ref{fig:slopes} shows the evolution of the $\dot{M}-M_\star$ slope over time for the best-fit synthetic populations. 
In this case, we did not add neither the observational spread nor the age spread to synthetic populations.
The figure illustrates that at the time at which the synthetic populations best reproduce Lupus, indicated by the blue dot for the viscous model and by the orange square for the MHD winds model, the predicted slopes (blue and orange, respectively) are consistent with the observational values (grey lines).
The viscous slope starts from $\lambda_{\dot{M}_0}\sim 1.1$, which, since $\lambda_{M_0} \sim 1.8$, corresponds to the regime where discs around low-mass stars evolve more rapidly, resulting in a steepening of the slope with time \citep{Somigliana2024} in the first 2 Myr.
In contrast, the MHD winds model predicts $\lambda_{M_0}-\lambda_{\dot{M}_0} \sim -0.5$, which causes discs around massive stars to evolve faster, resulting in a flattening of the slope.
At later stages, low-number statistics, particularly for the viscous-photoevaporative framework, which has a lower fraction of survivors (see Figure \ref{fig:5_Myr})), lead to a non-monotonic evolution of the slope.

\section{Viscous model at $\langle t \rangle  = 2$ Myr}
\label{viscous_model_2_Myr}

\begin{figure}[ht!]
    \centering
    \includegraphics[width=\linewidth]{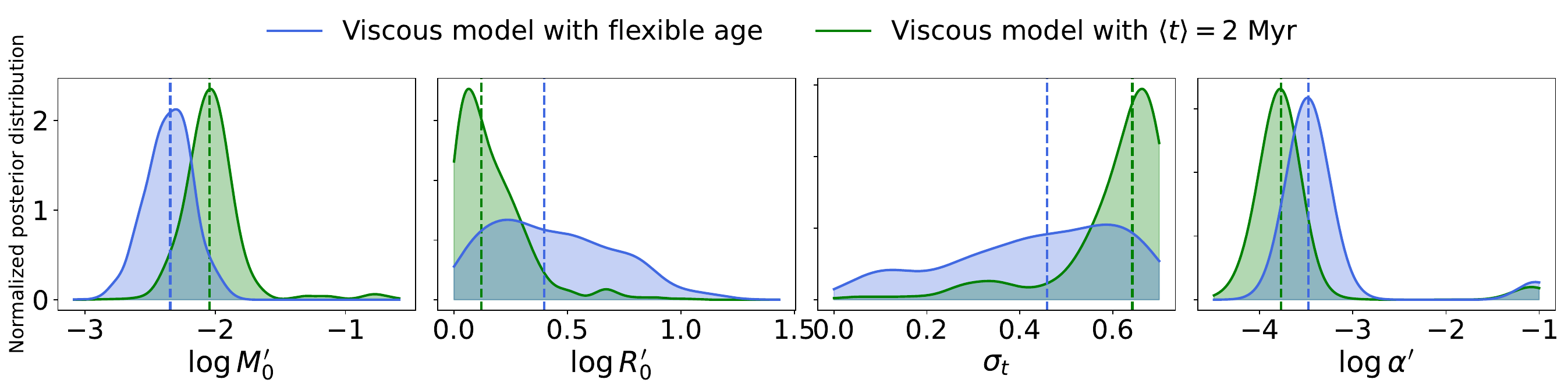} 
     
    \caption{\small{Comparison between the 1D posterior distribution of four parameters of the viscous-photoevaporative model, $\log M_0'$, $\log R_0'$, $\sigma_t$ and $\log \alpha'$, obtained applying \texttt{DEMOS} to the Lupus demographics.
    Blue curves show the distribution in when the mean age is treated as a free parameters, while green curves show the case when $\langle t \rangle$ is set at 2 Myr.
    Vertical dashed lines trace the medians of the distributions.}}
    \label{fig:comparison_fixed_age}
\end{figure}

By setting the mean age to 2 Myr, we obtain different posterior distributions of the model parameters.
The differences are more pronounced in the viscous-photoevaporative model, since its predicted age is not compatible with 2 Myr.
Figure \ref{fig:comparison_fixed_age} shows the 1D posterior distribution of four parameters of the viscous model when the mean age is included among the free parameters (blue) and when it is fixed to 2 Myr (green).
The figure highlights that, when $\langle t \rangle$  = 2 Myr, large initial masses are favoured.
In contrast, synthetic populations show a higher compatibility when the typical initial truncation radius, for a disc around a $0.3\ \text{M}_\odot$ star, is only 1.5-2 au.
Moreover, the $\log R_0'$ distribution is much narrower in the $\langle t \rangle$ = 2 Myr case, hinting that populations with $R_0 (M_\star = 0.3 \ \text{M}_\odot) > $3 au are not compatible with the data.
This result, combined with the preference for $\log \alpha' \sim 10^{-4}$, show that, if $\langle t \rangle = 2$ Myr, Lupus demographics are best reproduced by disc populations where discs are initially more compact and denser and less prone to viscous spreading.
This corresponds to the regime where the dominant evolution pathway, driven by internal photoevaporation, is outside-in dispersal \citep{Malanga2025}.
Indeed, the majority of synthetic discs, generated from these model parameters, undergoes this type of evolution.
This result recovers the findings of \cite{Tabone2025}.
In addition, the fit with fixed mean age shows a strong preference for a large age scatter, compatible with the upper limit of our prior, which is 0.7 Myr.
This is because the viscous model cannot reproduce observed discs with $\langle M/\dot{M}\rangle \sim 1 $Myr, unless these discs are much younger than the mean age (see Figure \ref{fig:age_scatter} for a comparison).

\section{Gravitational instability}
\label{GR}
We checked if the discs obtained with the initial conditions inferred by \texttt{DEMOS} are prone to be gravitationally unstable.
To do so, we computed the Toomre parameter $Q$ \citep{Toomre1964} defined as
\begin{equation}
    Q(r) = \frac{c_\text{s} \Omega_\text{k}}{\pi G \Sigma} = 2 \frac{H}{r} \frac{M_\star}{M} \frac{R_0}{r} \exp\left(\frac{r}{R_0}\right) \ ,
\end{equation}
where $c_\text{s}$ is the speed of sound and $\Omega_\text{k}$ the Keplerian angular velocity. 
For each disc, we computed the minimum $Q$, which occurs at $r = 3/4 R_0$, and we evaluated the fraction of discs with $Q_\text{min} \lesssim 1$, which is the condition for gravitational instability.
We obtain that with the initial condition inferred on Lupus with the viscous-photoevaporative model none of the discs are gravitationally unstable at $t = 0$ Myr, at the beginning of the disc phase.
For the MHD winds model, the fraction increases to $1.2 \%$.
This suggests, generally, that gravitational instability can be neglected in this phase of planet formation.
In addition, about 3/4 of the gravitationally unstable discs have $M_\star>1$ M$_\odot$, suggesting this phenomenon should primarily affect discs around intermediate and massive stars.

\rev{
\section{The effect of fixed spreads and a wider prior on $\log R_0'$}
\label{test_spread}

As Figures \ref{fig:Cornerplot_viscous} and \ref{fig:Cornerplot_MHD} show, some parameters are poorly constrained by the models.
In particular, most spreads of the initial conditions and $\log f_M$ for the MHD winds model, show distributions that are similar to the adopted priors or are pushed towards one of the boundaries of the prior range.
Therefore, for both models, we performed a fit where we fixed all spreads and $\log f_M$ to the median values obtained in the previous fit, and we checked whether the distribution of the remaining parameters became more precise.
For both models, we find that the posterior distribution of the remaining free parameters is almost unchanged (see Figure \ref{fig:mhd_spreads}).
\begin{figure}[ht!]
    \centering
    \includegraphics[width=0.8\linewidth]{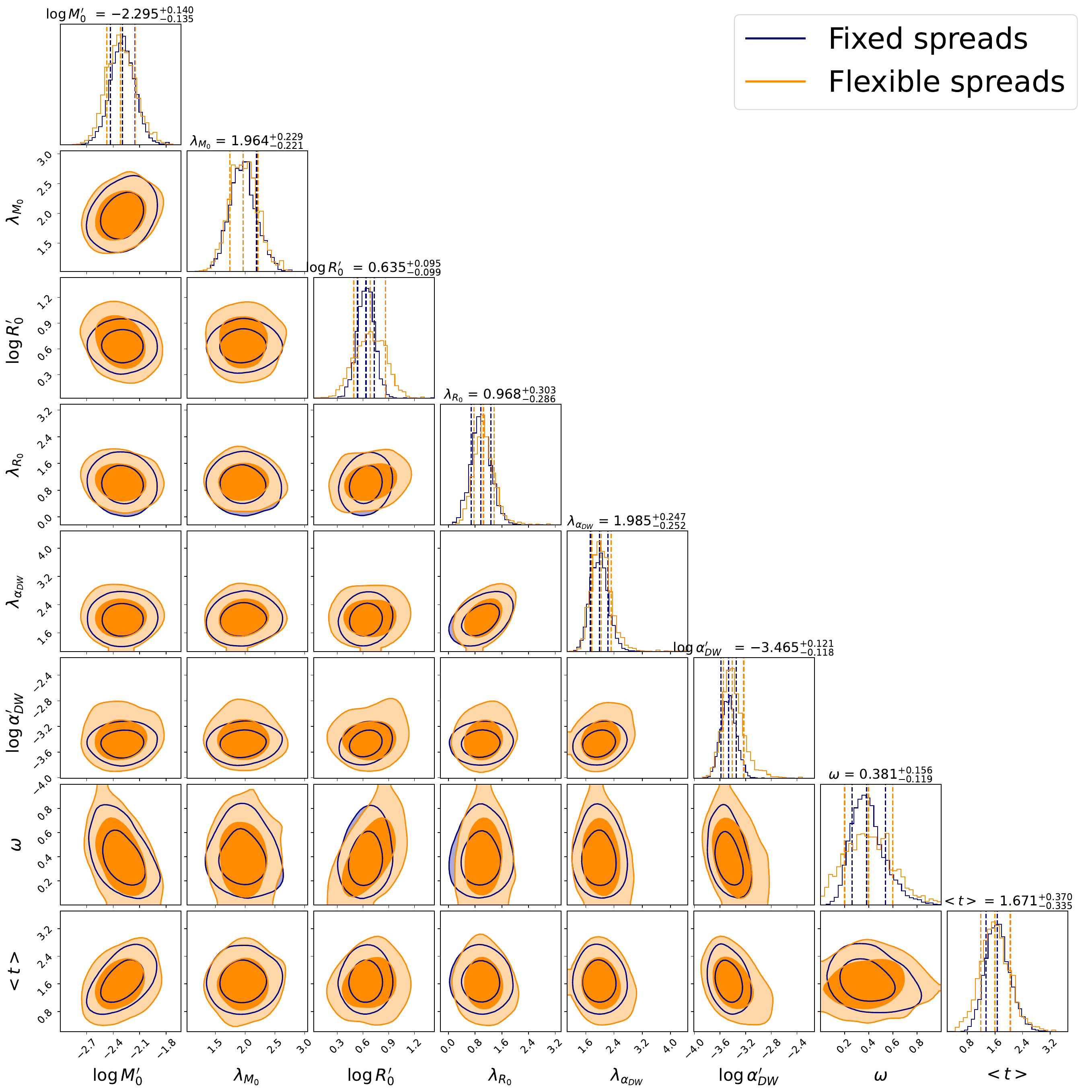} 
     
    \caption{\small{Comparison of the fit to Lupus using the MHD winds model with fixed (blue) and flexible (orange) spreads of the initial conditions. The latter corresponds to the model shown in Figure \ref{fig:Cornerplot_MHD}.
    Dashed vertical lines in the histograms indicate the 16$^\text{th}$, 50$^\text{th}$ and 84$^\text{th}$ percentiles, while numbers indicate the median and difference with respect to the 84$^\text{th}$ and 16$^\text{th}$ percentile of the case where spreads are fixed.
    Contours indicate the 1$\sigma$ and 2$\sigma$ confidence intervals.
}}
    \label{fig:mhd_spreads}
\end{figure}
Moreover, for both models, the Bayesian evidence of the fits with fixed spreads is not significantly higher than that of the corresponding fits with flexible spreads, with values of $\sim -256$ for the MHD winds model and $\sim -256$ for the viscous-photoevaporative model (which where $\sim -258$ and $\sim -255$, respectively, in the fits where spreads were not fixed), despite having five fewer free parameters. 

We also performed a fit with the viscous-photoevaporative model considering a wider prior for $\log R_0'$, in order to check whether its posterior distribution was artificially constrained by the narrow prior.
Therefore we chose $\mathcal{U}[-1,3]$ instead of $\mathcal{U}[0,2]$ of the previous fit.
The obtained 1D posterior distribution of $\log R_0'$ is shown in Figure \ref{fig:R_0_large_prior}, where it is compared to the one obtained previously.
\begin{figure}[ht!]
    \centering
    \includegraphics[width=0.5\linewidth]{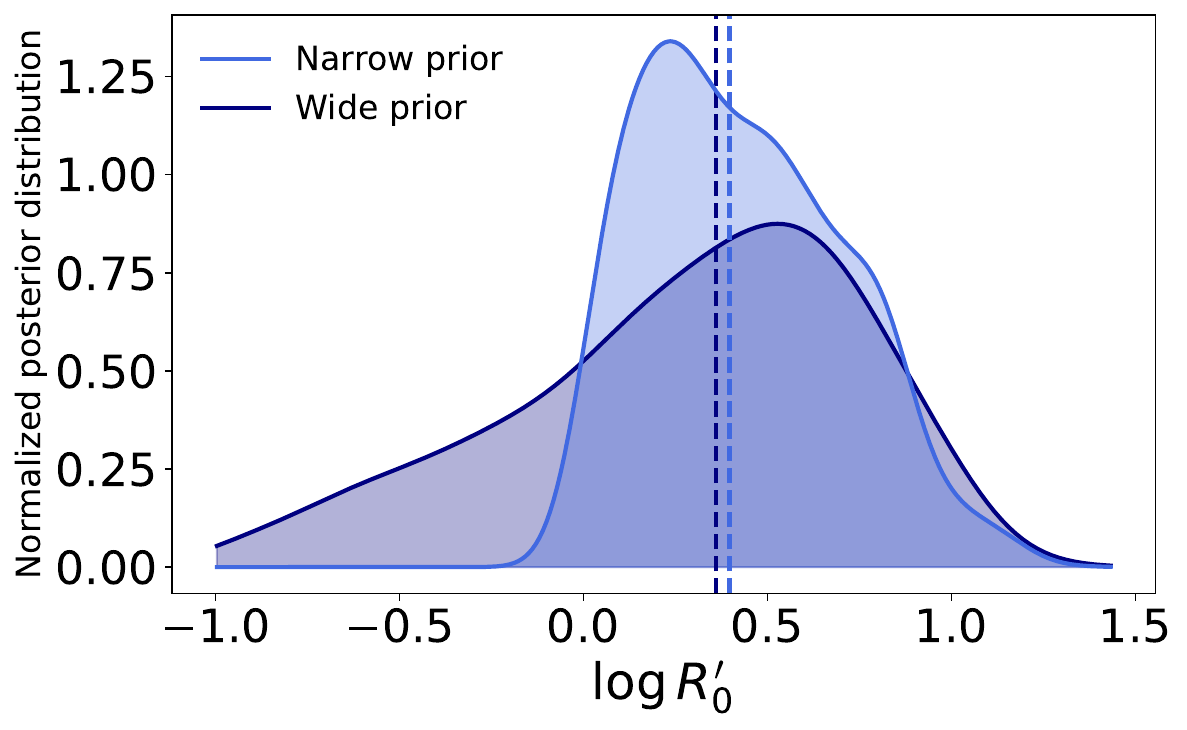} 
     
    \caption{\small{Comparison between the 1D posterior distribution of $\log R_0'$ with prior $\mathcal{U}[-1,3]$ (wide, dark blue) and $\mathcal{U}[0,2]$ (narrow, light blue), which is the same as Figure \ref{fig:Cornerplot_viscous}.
    Dashed vertical lines indicate the median of the distributions.
}}
    \label{fig:R_0_large_prior}
\end{figure}
The Figure shows that, although the wider prior allows a better constraint on the posterior distribution, where it is not pushed towards the lower limit, the medians of the two distribution are surprisingly close.
This suggests that, despite being narrow, the $\mathcal{U}[0,2]$ prior was not artificially constraining the initial disc size.
However, allowing for a wider prior showed us that solutions with initial sizes $R_0< 1 $au can be fairly compatible with the data.
Nevertheless, these solutions should be treated with caution, due to the limited resolution of the \texttt{Diskpop} grid at $r<1$ au and due to the large fraction of upper limits on disc sizes in the data.
Measurements of disc sizes for compact sources will therefore be crucial for further investigating the compatibility of extremely compact disc models with the observations.
In addition, we performed a simulation with the MHD winds model, where we widened the prior of $\log f_M$ to $\mathcal{U}[-2,2]$.  As expected, we found that $f_M$ is unconstrained for values much smaller than unity, since it enters the governing equations only through the factor (1+$f_M$).
}
\end{appendix}

\end{document}